\documentclass[preprint,12pt]{elsarticle}

\usepackage{amssymb}
\usepackage[T1]{fontenc}
\usepackage{caption}
\usepackage{subcaption}
\usepackage{float}
\usepackage{array}
\usepackage{tabularray}
\usepackage{graphicx}
\usepackage{amsmath}
\usepackage{longtable}
\usepackage{url}
\usepackage{xcolor}

\journal{Computer Networks}

\begin{document}

\begin{frontmatter}



\title{Diver-Robot Communication Dataset for Underwater Hand Gesture Recognition}

\author[FER]{Igor Kvasić\corref{cor1}}\ead{igor.kvasic@fer.hr}
\author[ABI]{Derek Orbaugh Antillon\corref{cor1}}\ead{dorb476@aucklanduni.ac.nz}
\author[FER]{\DJ ula Na\dj}
\author[ABI]{Christopher Walker}
\author[ABI]{Iain Anderson}
\author[FER]{Nikola Mišković}

\cortext[cor1]{Corresponding authors.}

\newcommand{\orcidauthorA}{0000-0003-4705-1852} 
\newcommand{\orcidauthorB}{0000-0002-4690-246X} 
\newcommand{\orcidauthorC}{0000-0001-9717-5218} 
\newcommand{\orcidauthorD}{0000-0003-2867-2944} 
\newcommand{\orcidauthorE}{0000-0003-4698-7317} 
\newcommand{\orcidauthorF}{0000-0003-1474-4126} 


 \affiliation[FER]{organization={Laboratory for Underwater Systems and Technologies, Faculty of Electrical Engineering and Computing, University of Zagreb},
            addressline={Miramarska 20},
            city={Zagreb},
            postcode={10000},
            country={Croatia}}
 \affiliation[ABI]{organization={Auckland Bioengineering Institute, The University of Auckland},
            addressline={6/70 Symonds Street, Grafton},
            city={Auckland},
            postcode={1010},
             country={New Zealand}}

\begin{abstract}
{In this paper, we present a dataset of diving gesture images used for human-robot interaction underwater. \textcolor{black}{By offering this open access dataset, the paper aims at investigating the potential of using visual detection of diving gestures from an autonomous underwater vehicle (AUV) as a form of communication with a human diver. In addition to the image recording, the same dataset was recorded using a smart gesture recognition glove.} The glove uses elastomer sensors and on-board processing to determine the selected gesture and transmit the command associated with the gesture to the AUV via acoustics. Although this method can be used under different visibility conditions and even without line of sight, it introduces a communication delay required for the acoustic transmission of the gesture command. To compare efficiency, the glove was equipped with visual markers proposed in a gesture-based language called CADDIAN and recorded with an underwater camera in parallel to the glove's onboard recognition process. The dataset contains over 30,000 underwater frames of nearly 900 individual gestures annotated in corresponding snippet folders. The dataset was recorded in a balanced ratio with five different divers in sea and five different divers in pool conditions, with gestures recorded at 1, 2 and 3 metres from the camera. The glove gesture recognition statistics are reported in terms of average diver reaction time, average time taken to perform a gesture, recognition success rate, transmission times and more. The dataset presented should provide a good baseline for comparing the performance of state of the art visual diving gesture recognition techniques under different visibility conditions.}
\end{abstract}




\end{frontmatter}



\section{Introduction} \label{introduction}
Underwater human-robot interaction is a challenging field of research due to several factors unique to the underwater environment. Despite the challenges, researchers have recently been making significant progress in underwater human-robot interaction, developing advanced control algorithms, communication systems, and sensory technologies that improve the capabilities of underwater robots and enhance their collaboration potential with human divers \cite{uhri_overview}. Some of these challenges and their effects are being addressed in ONR project "ROADMAP -  Robot-Aided Diver Navigation in Mapped Environments" \cite{roadmap}, which aims at enhancing diver capabilities and safety using underwater robots as diving buddy counterparts \cite{remote_experiments}. The robot, using an assortment of industrial grade sensors, can assist the diver in a variety of underwater missions aiding his localization, navigation, communication and perception capabilities \cite{diver_assistance}. When humans interact with robots underwater, safety becomes of utmost concern \cite{safetyreq, human_factor}. Human divers face continuous risks and if a robot malfunctions or behaves unexpectedly, it can lead to life-threatening situations \cite{auvdivingrisks}. As the underwater environment may not always provide excellent visibility conditions, it is crucial for the diver to have a reliable means of communicating and interacting with the robot.
The underwater environment is inhospitable to both humans and robots, but some of the risks associated with demanding diving missions can be mitigated using recent advances in technology. Underwater communication is much more challenging than communication in air. Water significantly attenuates radio signals, making wireless communication unreliable \cite{underwater_communication}. Acoustics are commonly used for communication, but it faces issues such as significantly limited bandwidth, signal distortion and acoustic noise, which all can hinder real-time and precise interactions between humans and robots \cite{acoustic_comms}. Furthermore, water is often turbid and lacks light at higher depths, leading to reduced visibility even in the clearest of conditions \cite{WRIGHT1995}. Human-robot interaction in general relies heavily on visual feedback, which poses a significant challenge when moved underwater. This limitation affects both remote operation and the ability of autonomous robots to perceive and respond to human gestures and cues. In both remote-controlled and autonomous operations, there is often a significant communication latency between the human operator and the underwater robot. This delay can make real-time control and responsiveness more difficult, especially in critical situations.
On the other hand, humans are highly dexterous, allowing us to perform intricate tasks with ease, but it is also common for divers to use hand gestures as means of underwater communication \cite{gesture_communication}. Previous studies such as \cite{caddy-general} research the possibility of using an underwater tablet for communication, but diving gloves significantly reduce the dexterity and touch screen possibilities, and using a dedicated pen and carrying the tablet proved to be too cumbersome for the diver. Within project cited in \cite{caddian}, authors proposed using standard and well established hand gestures widely adopted among divers, such as PADI \cite{padi} and CMAS \cite{CMAS} gestures, as a natural form of communication between divers, to be used for conveying different messages to an autonomous robot. A complete diver-robot communication language called the CADDIAN was composed in \cite{caddian2}, with detailed semantics and instructions for performing the gestures. The paper proposes an initial alphabet of 40 context dependent commands defined by syntax, semantics and communication protocol used to communicate with the AUV. \textcolor{black}{ Authors from the same research groups proceed to publish the "CADDY" underwater stereo image dataset \cite{CADDY-dataset} containing nearly 10K image pairs collected over several field trials in various environmental conditions.} The diving glove used in their experiment was equipped with vinyl strips in different colors on each finger and rectangle and circle signs on the palm and back respectively in order to enhance visual recognition between gestures \cite{underwater_image_processing}. It is exactly this type of glove and gesture vocabulary that is replicated in this paper, in addition to the glove containing smart gesture recognition sensors. 

\textcolor{black}{The implementation of a gesture-based communication protocol between the diver and the robot introduces several network implications, particularly in scenarios where the robot acts as an intermediary relay to the surface via an acoustic link. In the proposed scenario where the AUV acts as a diving buddy counterpart, it has the task of maneuvering in the vicinity of the diver and interpreting diving gestures as a natural form of communication for the diver. Once this communication is established, the AUV serves as an acoustic communication relay link to transmit or receive messages from a surface robot, offshore or onshore command center, or other underwater agents. The robot also plays a key role as a navigation aid for the diver that can transmit the current position or receive new mission waypoints.}

\textcolor{black}{While the CADDY dataset focuses on a stereo imagery approach investigating 3D perception methods and is part of a broader human-robot tracking and interaction scenario research, our work focuses on using single camera acquisition for exploiting widely used image classification methods for communication purposes. To our knowledge there are no similar publicly available datasets and we hope that our efforts will help bridging the gap between land and underwater perception frameworks originating from data scarcity.}

\textcolor{black}{Considering all the previous research done in the field of gesture-based underwater communication we can conclude that it is showing to be a promising approach for underwater human-robot interaction. The work presented in this paper relies on that hypothesis and aims to build on it providing the following contributions:
\begin{itemize}
    \item introducing an open, publicly available dataset of diving gesture images for visual machine recognition
    \item providing gesture detection results using a gesture recognition glove with onboard processing and acoustic message transmission
    \item providing a detailed semantic and syntactic gesture definition and description of the collected data with respect to the environment and conditions
    \item presenting relevant quantitative metrics for benchmarking machine vision gesture recognition methods"
\end{itemize}
}

\section{Methodology and sensor setup}

\begin{figure} [H]  
\begin{center}
\includegraphics[width=10.5 cm]{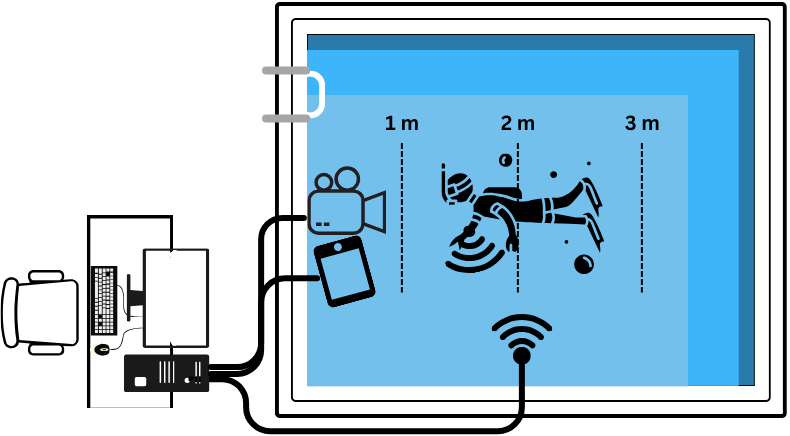}
\caption{Experiment overview: data acquisition PC and control center on the left, underwater camera and tablet placed on the waterside facing the diver, acoustic modems mounted on the sidewalls and the diver performing gestures using the gesture recognition glove at 1, 2 and 3 meter marks.} 
\label{fig:overview}
\end{center}
\end{figure}

In order to simultaneously record both the visual gesture data required to create the dataset and the sensory data from the glove, the experiment is set up as follows. An underwater camera is placed on the waterside wall, in both pool or sea scenario, facing the diver in front. One, two and three meter marks are placed on the bottom in front of the camera so that the diver can easily position himself at the correct distance from the camera. An underwater tablet has been placed next to the camera to communicate with the diver during the experiment and for displaying the gesture to be performed next. The gesture recognition glove processes the data directly on board and, once a gesture is successfully detected, sends an associated glove command via a compact acoustic transmitter integrated into the glove. To capture that acoustic message, an acoustic receiver was placed laterally to the diver-camera line, at about three meters away. All this is connected to a poolside PC that does data acquisition but also serves as a command center for the experiment progress. An overview of the experiment is visualised in Figure \ref{fig:overview}.

\subsection{Gesture Recognition Glove}

The diver's glove investigated in \cite{CADDY-dataset} and replicated in this paper has been customized by incorporating colored vinyl strips on each finger and custom shapes marking the palm and back of the hand. A 5 cm diameter circle was placed on the palm of the glove, while on the back a 5 cm sided square, both made with a 1 cm white border. The primary purpose of the visual markers is to enhance visual characteristics of the glove for more effective feature extraction in tasks related to classification. Completely black gloves were found difficult to differentiate from the rest of the diver's body, even to the human eye. In the context of developing methods for visual classification of images and gesture detection, color strips serve as a foundation of exploiting color information as an additional visual feature.

\begin{figure}[H]
\begin{subfigure}{\textwidth}
  \centering
  \includegraphics[width=0.6\textwidth]{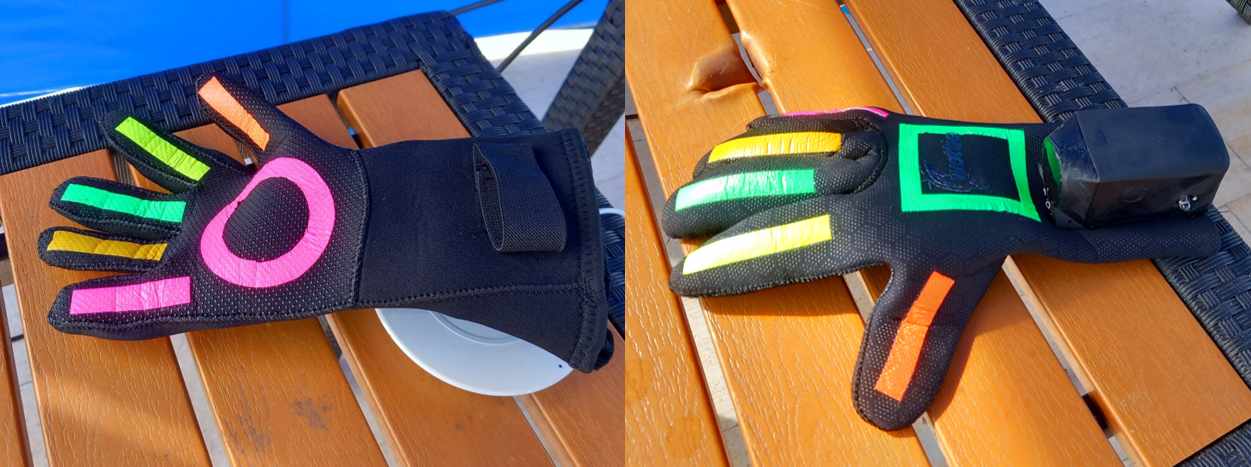}
  \label{nasa}
\end{subfigure}
\begin{subfigure}{\textwidth}
    \centering
  \includegraphics[width=0.6\textwidth]{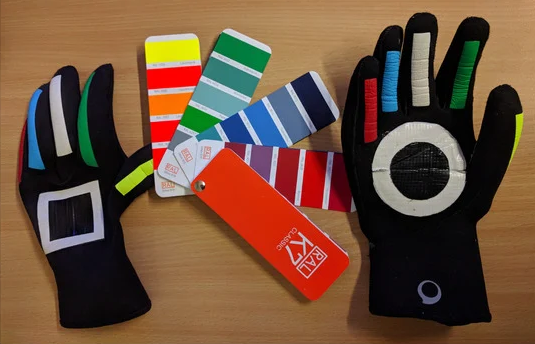}  
  \label{caddy}
\end{subfigure}
\caption{Gesture recognition glove used in collecting the dataset, modified with color strips replicating the CADDIAN diver gloves (above) and the original CADDIAN \cite{CADDY-dataset} glove (below) against a RAL-K7 chart.}
\label{fig:gloves}
\end{figure}

In parallel to the visual recording of the diver's hand during performing gestures, the hand movement was recorded using the sensors integrated in the diving glove \cite{glove}. Wearable stretch and inertial sensors are seamlessly integrated into the diver's glove to enable the recognition of static and dynamic gestures made by the diver. The sensors, dielectric elastomers, are specially designed for underwater use and are embedded within the diver's glove where they capture the movement of each finger \cite{elastomers}. In addition to elastotomer sensors placed on each finger, the glove features an inertial measurement unit (IMU) sensor placed on the back of the hand in order to provide the wrist orientation and movement. Using roll, pitch, yaw and acceleration from the IMU enables distinguishing between similar gestures done in different hand orientations (e.g. stop sign or level sign) or incorporating dynamic gestures that use hand movement. The sensor readings were interpreted directly on the glove electronics to detect the gesture with the highest probability of being selected. When a gesture is recognised correctly, the diver receives feedback through a combination of haptic vibrations and LED signals. Since every user has different hand anatomy and range of motion, an initial sensor range calibration is conducted for each user upon powering the glove. An auto-calibration procedure is employed during the whole operation to enhance robustness of the recognition.
Raw data of the glove sensor readings were stored directly on the glove electronics memory and can be downloaded over Bluetooth after the mission for post-processing. 
Once a gesture is successfully recognised, the gesture information is encoded in a message that is sent acoustically to a receiver connected to the poolside PC. The glove is equipped with an acoustical modem transmitting in the 24-32 kHz frequency range and is able achieve an acoustic throughput of up to 463 bits/s. Although not large, the throughput is enough to periodically transmit relevant data and gesture commands. In a diver-robot interaction scenario proposed in Section \ref{introduction}, the glove sends the acoustic data directly to the diving "buddy" AUV. For this data collection purposes, the receiving end was an acoustical receiver connected to the poolside PC via a serial connection. 

\subsection{Camera Setup}

The SwimPro Wallcam camera ~\cite{swimpro}, which is commonly used to create ad hoc video recording pool setups for swimming competitions or for detailed swimming analyses of top athletes, was used to capture the image dataset. The camera offers recording options at 30 or 60 frames per second and features an image resolution of up to 2k (2560×1440). Both wireless and wired (Ethernet) configuration can be use to transmit the data, but in order to minimize the video transmission latency, the wired option was used. The camera features a nearly 100 degree wide horizontal field of view with proprietary Linear View Technology, which reduces the fisheye effect.

\begin{figure} [H]  
\begin{center}
\includegraphics[width=0.7\textwidth]{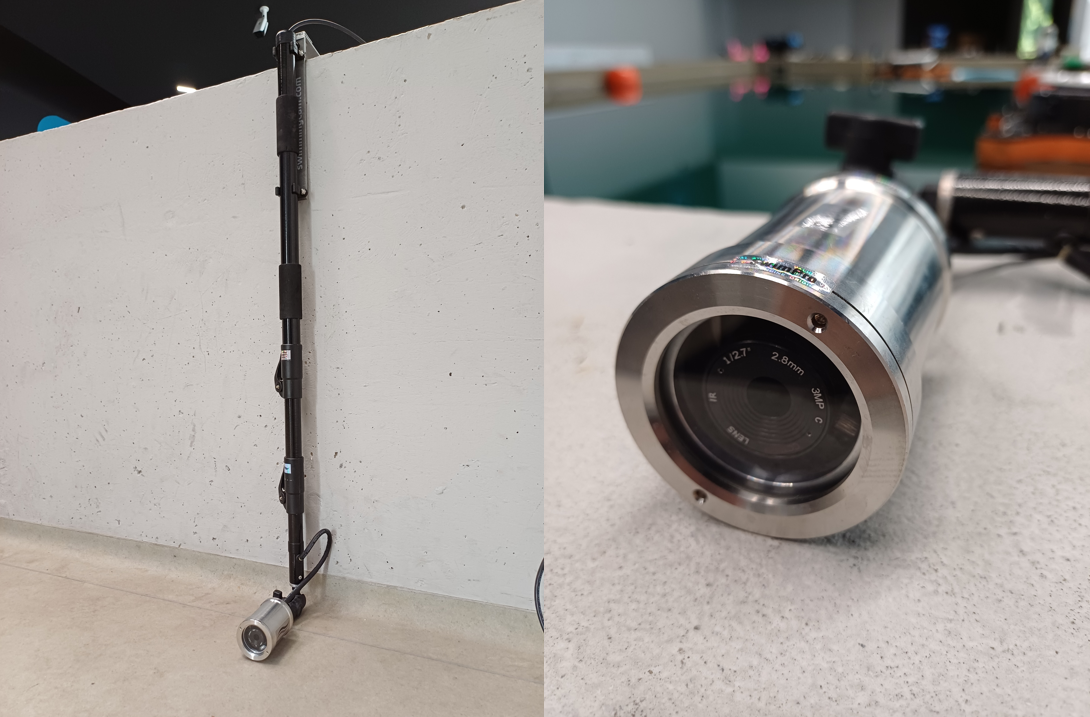}
\caption{Swimpro ~\cite{swimpro} camera used for recording underwater images on the telescopic pole for mounting on the pool sidewall.} 
\label{fig:cameras}
\end{center}
\end{figure}

The camera was connected to the PC and powered over an Ethernet connection. A custom Unity application described in subsection \ref{unity} was created to display and capture the footage. The camera was mounted on the pool or shoreline wall and set at 2 meters of depth. To position the diver at exactly 1, 2 and 3 meters from the camera, distance was measured and marked with yellow painted weights easily seen by the divers when underwater.

\subsection{Underwater Tablet}

In order to communicate with the diver from the surface once in the water and to coordinate the overall progress of the mission, an underwater tablet was used as a visual interface to the diver. The interface consists of a standard 10-inch tablet securely housed in a custom-built waterproof case designed to withstand pressure up to 50 meters of depth. To enable a connection to the tablet, a Bluetooth modem molded in epoxy resin is attached to the back of the waterproof housing, ensuring Bluetooth connectivity with the tablet without compromising the structural integrity of the housing. Although water significantly attenuates radio waves of such frequencies, the connection was stable as the two devices were positioned in close proximity to each other. Although the tablet has the option of utilising a modified inductive pen to preserve touchscreen functionality underwater, this was not used. Instead, visual communication via standard video gestures was used as two-way communication.

\begin{figure}
\begin{center}
\includegraphics[width=\textwidth]{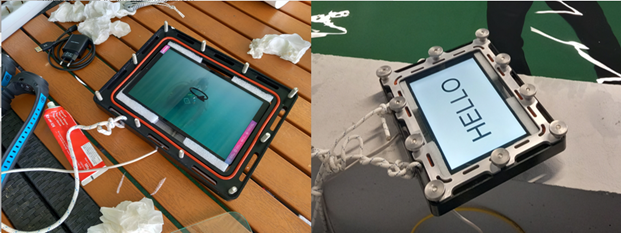} 
\caption{Tablet in the underwater casing used to interface with the diver once underwater and communicate the gesture sequence to be performed.} 
\label{fig:tablet}
\end{center}
\end{figure}

The tablet was running a custom Unity Android application communicating in serial with the poolside PC. During the mission, the diver was shown pictures with information about the gesture to be performed next to ensure the same sequence for all the divers, but also as a reminder them of the way each gesture is performed. Additionally, messages about the mission such as distance to the camera, gestures to be repeated or instructions to surface by the end of the mission  the diver. The tablet in the underwater casing with an example of a gesture instruction can be seen on Figure \ref{fig:tablet}.

\subsection{Unity Application} \label{unity}

\begin{figure}
\begin{center}
\includegraphics[width=0.8\textwidth]{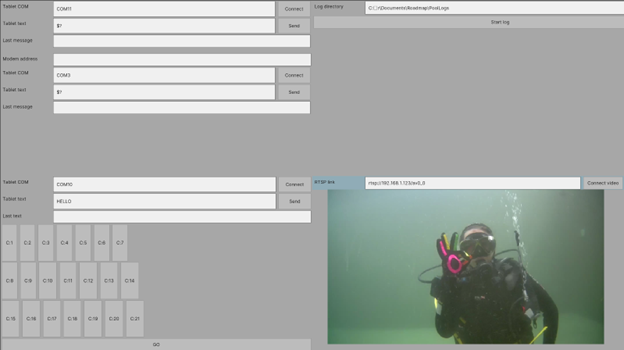} 
\caption{User interface of the Unity application for communicating with the diver underwater.} 
\label{fig:screenshot}
\end{center}
\end{figure}

To record the underwater video but also to achieve synchronised acquisition of all recorded sensors (glove, tablet video and acoustic modem) a custom application using Unity engine \cite{unity} was made. The application served as an interface to the topside operator during the mission and was divided in four sections as can be seen on Figure \ref{fig:screenshot}. The parameters for the acoustic modem connection such as baud rate, communication port and modem address are set in the upper left part of the screen. The messages received via the acoustic modem are also displayed in this area as soon as the glove has successfully recognized a gesture. For redundancy reasons, two receiving acoustic modems have been used at slightly offset positions, which has helped to reduce unsuccessful message reception due to multipath issues. The bottom left of the screen is used to select messages to be displayed to the diver on the underwater tablet. Predefined messages with instructions of gestures used during the mission were programmed into buttons for a more convenient use, as well as common messages for positioning at 1, 2 or 3 meters from the camera, surfacing or repeating a message. Custom messages can also be typed in a message field and are displayed as text on the tablet screen. The bottom right part of the screen was reserved for displaying the video feed from the underwater camera. Camera was connected over Ethernet and video was streamed using Real Time Streaming Protocol (RTSP). The top right of the screen was used for selecting the output file path and starting of stopping the logging of data. The output file was in .csv format including a precise timestamp followed by messages sent from the tablet and messages received from the acoustic modem. The video was stored separately in .mkv format along with frame timestamped data.

\section{Dataset}

\subsection{Underwater Gestures}

Underwater communication between diver buddies is essential for maintaining safety, coordinating actions, and sharing information during a dive \cite{gestures}. While there isn't a single universal standard for gesture communication among divers, some common hand signals are widely recognized and used in the diving community. These signals help divers convey important messages without relying on verbal communication, which can be challenging underwater due to the limited audibility and potential disturbance to marine life. 

One of most recognisable and standard hand signals commonly used between diver buddies is the OK sign: to signal that everything is okay, a diver typically forms a circle with the thumb and index finger while extending the other three fingers upward. To indicate the direction of ascent or descent for example, a diver will point their thumbs up finger upward or downward, respectively. To inform a buddy that they are out of air, a diver may simulate a "throat cut" motion across their neck or repeatedly tap their hand on their tank.
It's important to note that while these hand signals are widely understood, diver buddies should agree on specific signals before the dive to ensure clear communication and avoid misunderstandings. Similarly, when considering a gesture set to be used between a diver and a robot, dive or mission specific tasks should be addressed and assigned the most suitable gestures. Additionally, different organizations and training agencies may have their variations or additional signals, so divers should be aware of any specific communication protocols associated with their training and certification.

From the CADDIAN language set, which is heavily based on the most common organisation's protocols, the following eight were chosen for this experiment: "ear problem", "out of air", "out of breath", "general evacuation", "OK", "reserve" and "turn". Examples of the selected gestures being performed underwater together with each gesture semantics can be found in Table \ref{sve_geste}.
\noindent 
\begin{longtblr}[
    caption = {Set of eight main diving gestures used in the experiment with gesture meaning, semantics and instructions for performing the gesture on the left, and a diver performing the gesture underwater on the right.},
    label = {sve_geste},
]{
  colspec = {X[c,h]X[c]},
  stretch = 0,
  rowsep = 2pt,
  hlines = {1pt},
  vlines = {1pt},
  colsep = {1pt},
}

\textbf{Gesture} & \textbf{Image}\\*
{\textbf{Ear problem} \\ Semantics: I have an ear-related problem \\ Type: static gesture \\ Hands: R \\ Palm/back: back to camera \\ Fingers: 1R} & {\includegraphics[width=0.49\textwidth]{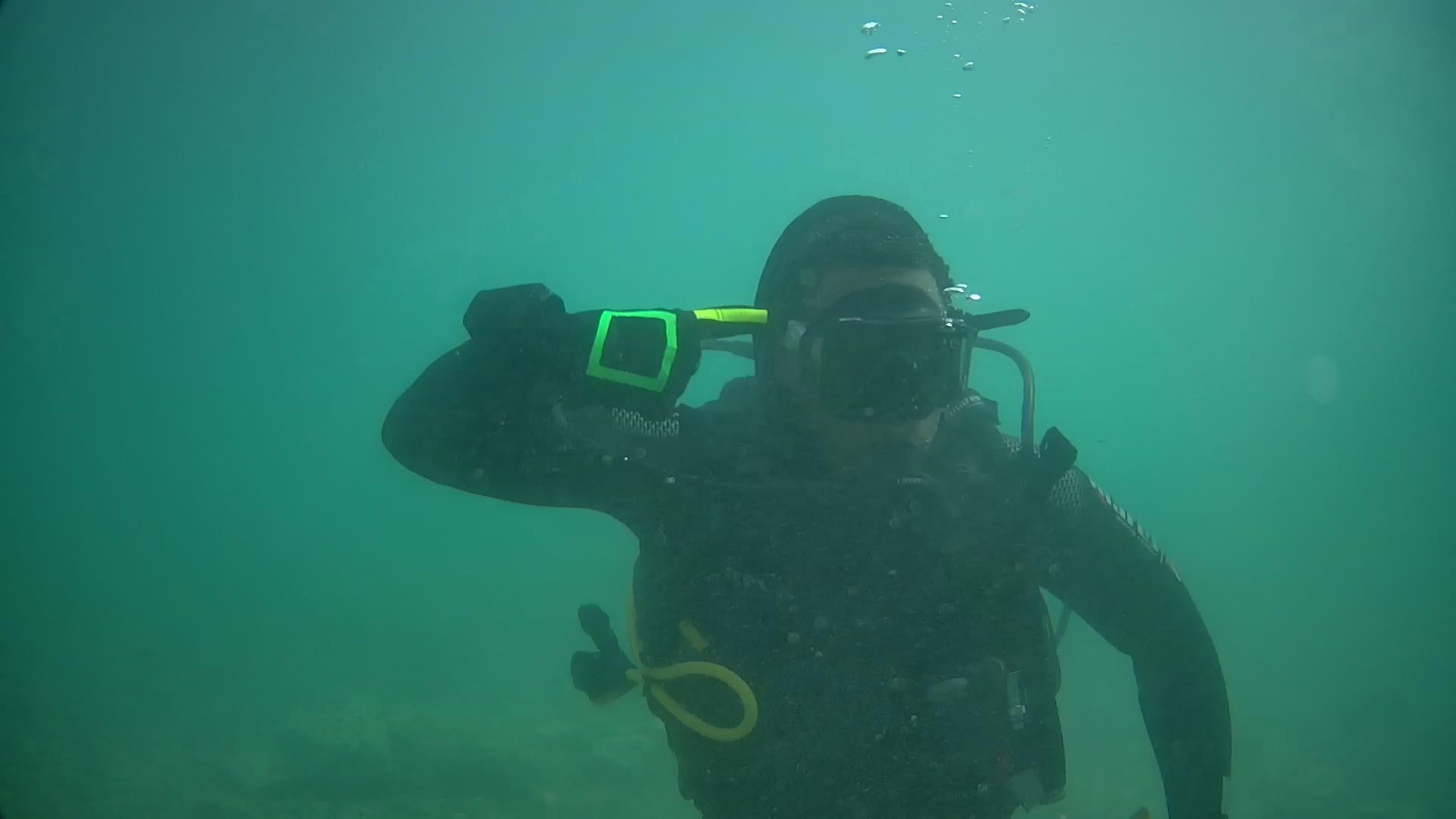}}\\
{\textbf{Out of air} \\ Semantics: I’m out of air \\ Type: static gesture \\ Hands: R \\ Palm/back: back slightly towards the camera \\ Fingers: 1R,2R,3R,4R,5R} & {\includegraphics[width=0.49\textwidth]{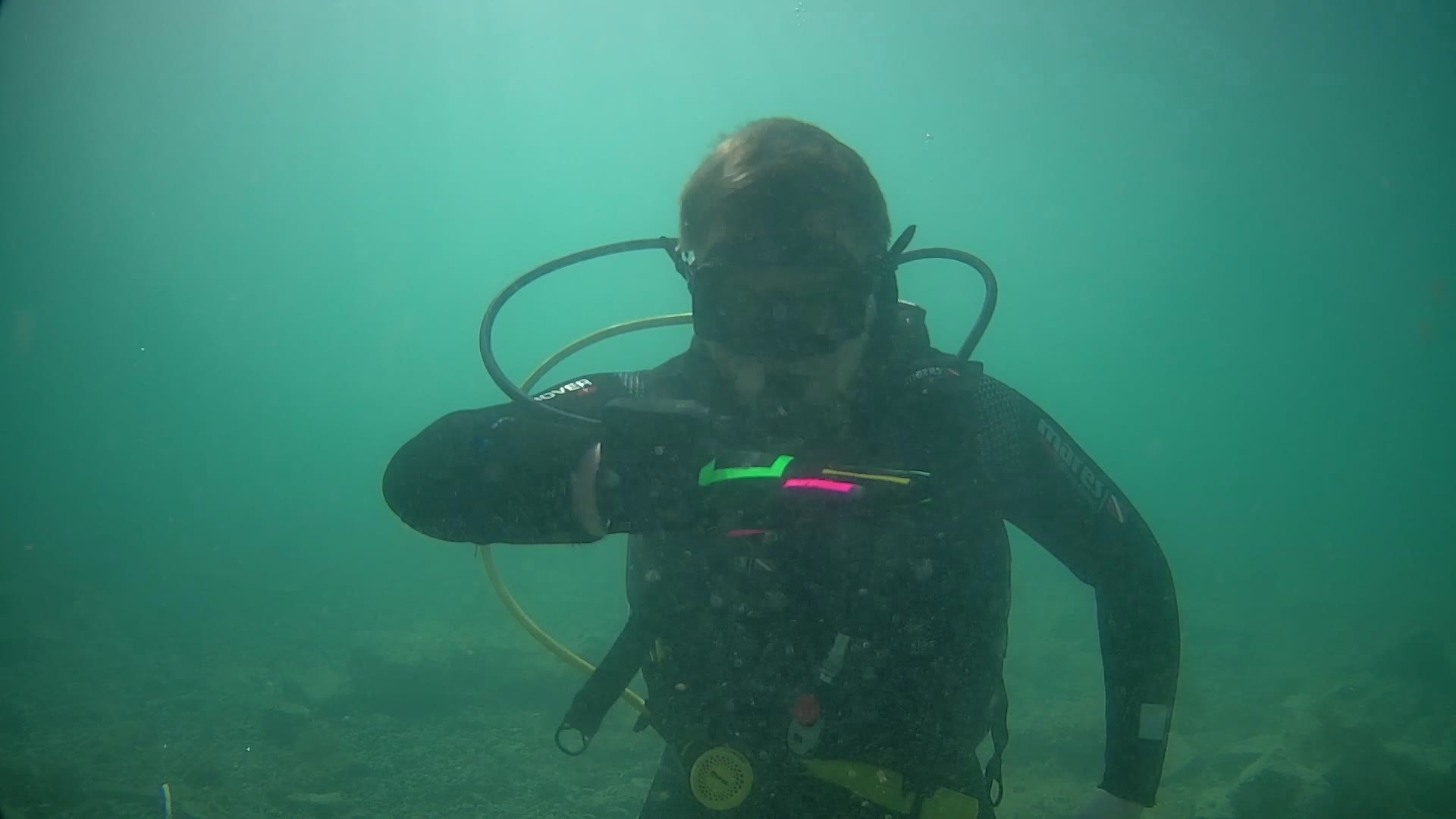}}\\
{\textbf{Out of breath} \\ Semantics: I’m out of breath \\ Type: static gesture \\ Hands: R \\ Palm/back: back to camera \\ Fingers: 1R,2R,3R,4R,5R} & {\includegraphics[width=0.49\textwidth]{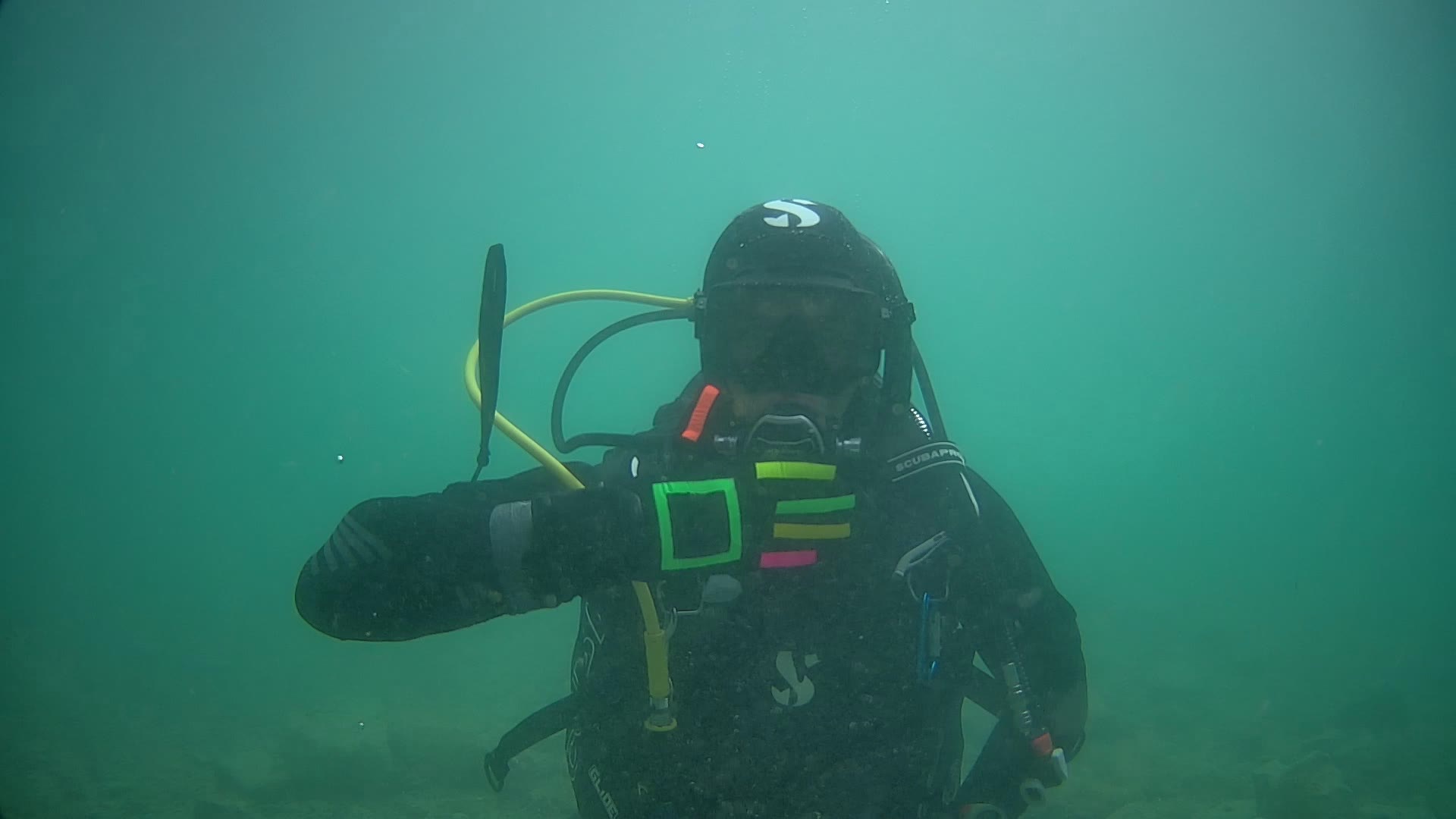}}\\
{\textbf{Danger} \\ Semantics: danger, something is wrong \\ Type: static gesture \\ Hands: R \\ Palm/back: back to camera \\ Fingers: fist} & {\includegraphics[width=0.49\textwidth]{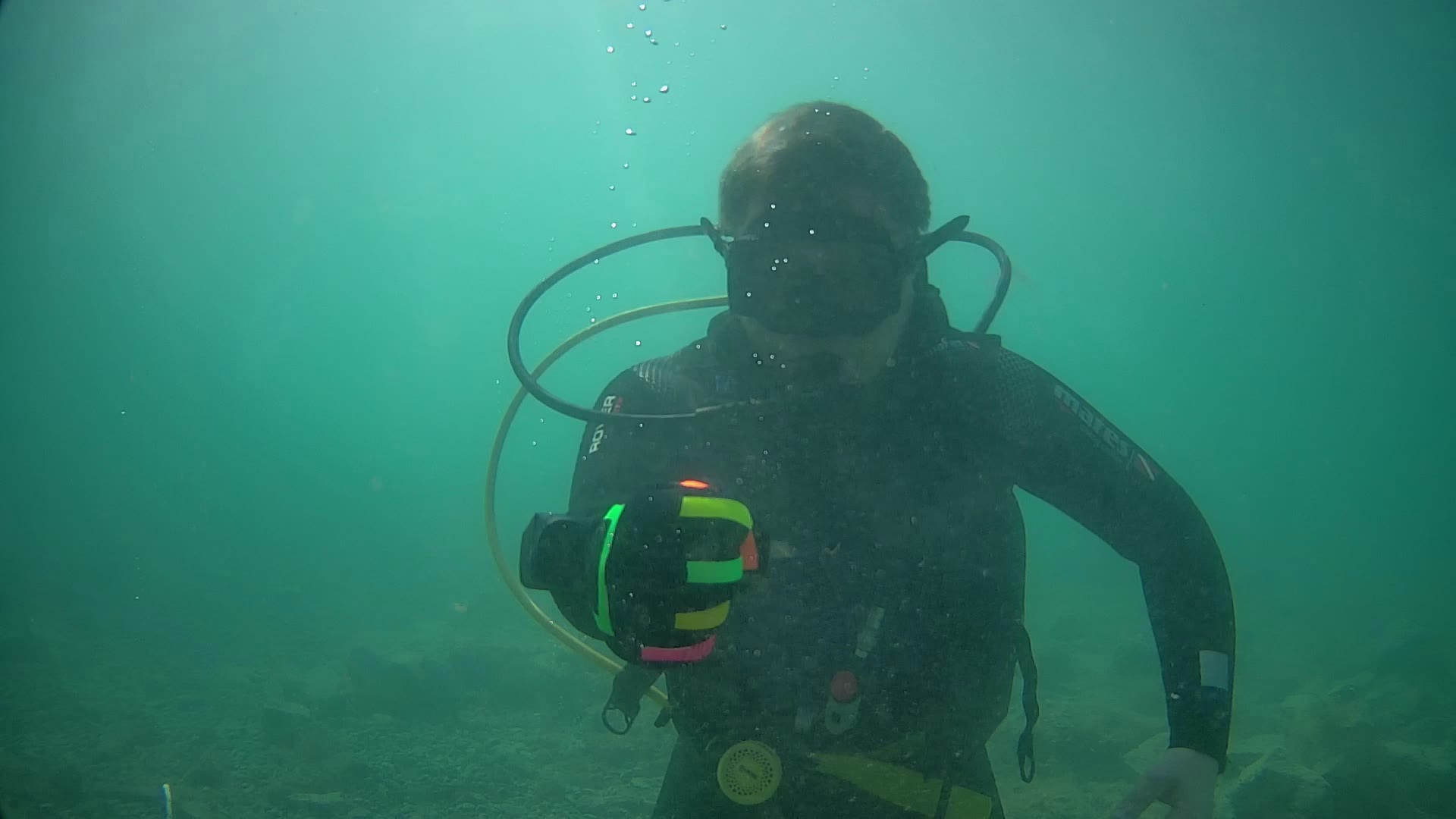}}\\
{\textbf{General evacuation} \\ Semantics: general evacuation \\ Type: static gesture \\ Hands: R \\ Palm/back: back to camera \\ Fingers: 1R,2R} & {\includegraphics[width=0.49\textwidth]{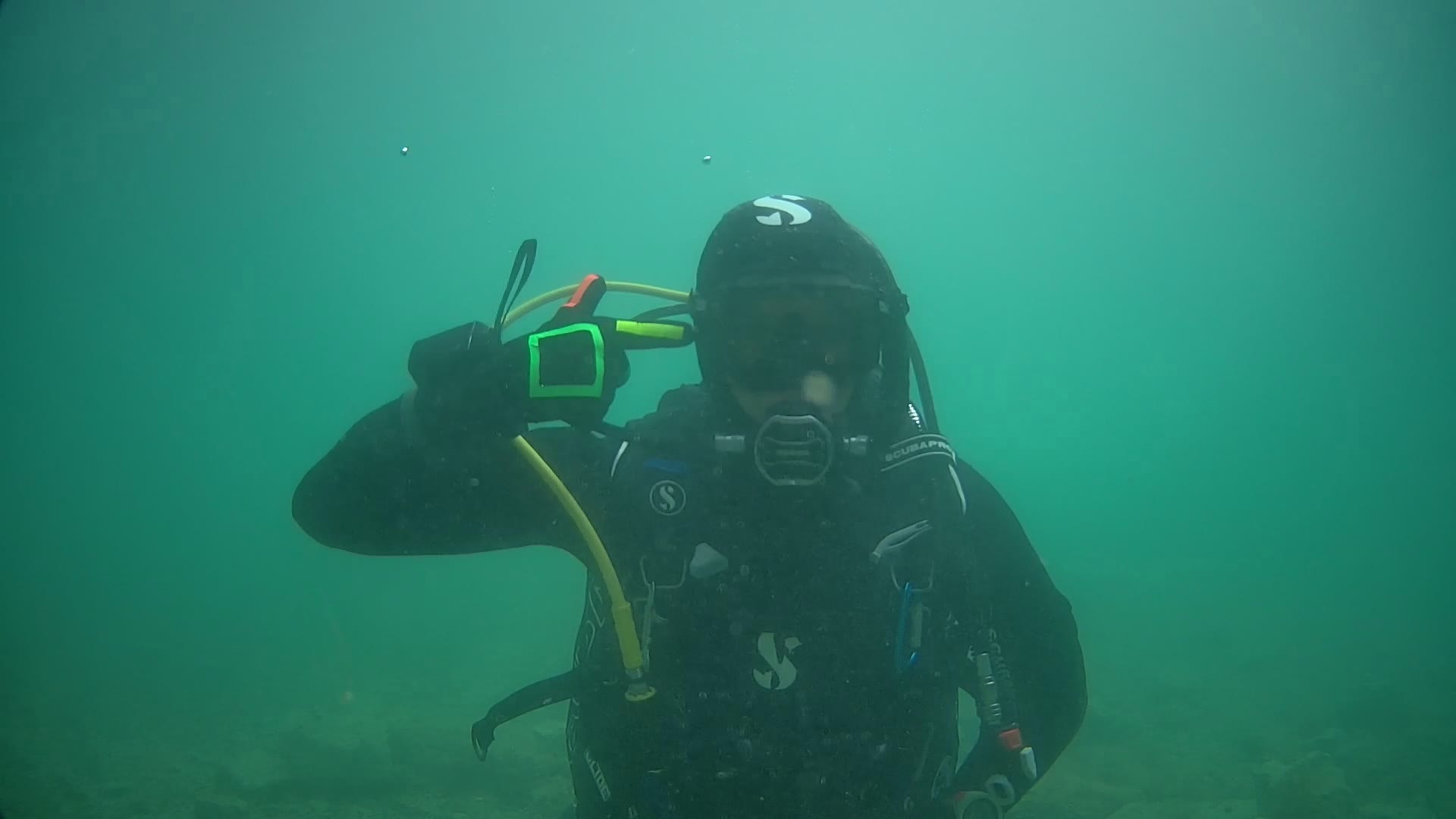}}\\
{\textbf{OK} \\ Semantics: ok [answer], let’s go \\ Type: static gesture \\ Hands: R \\ Palm/back: back to camera \\ Fingers: 3R, 4R, 5R AND 1R,2R bent and touching} & {\includegraphics[width=0.49\textwidth]{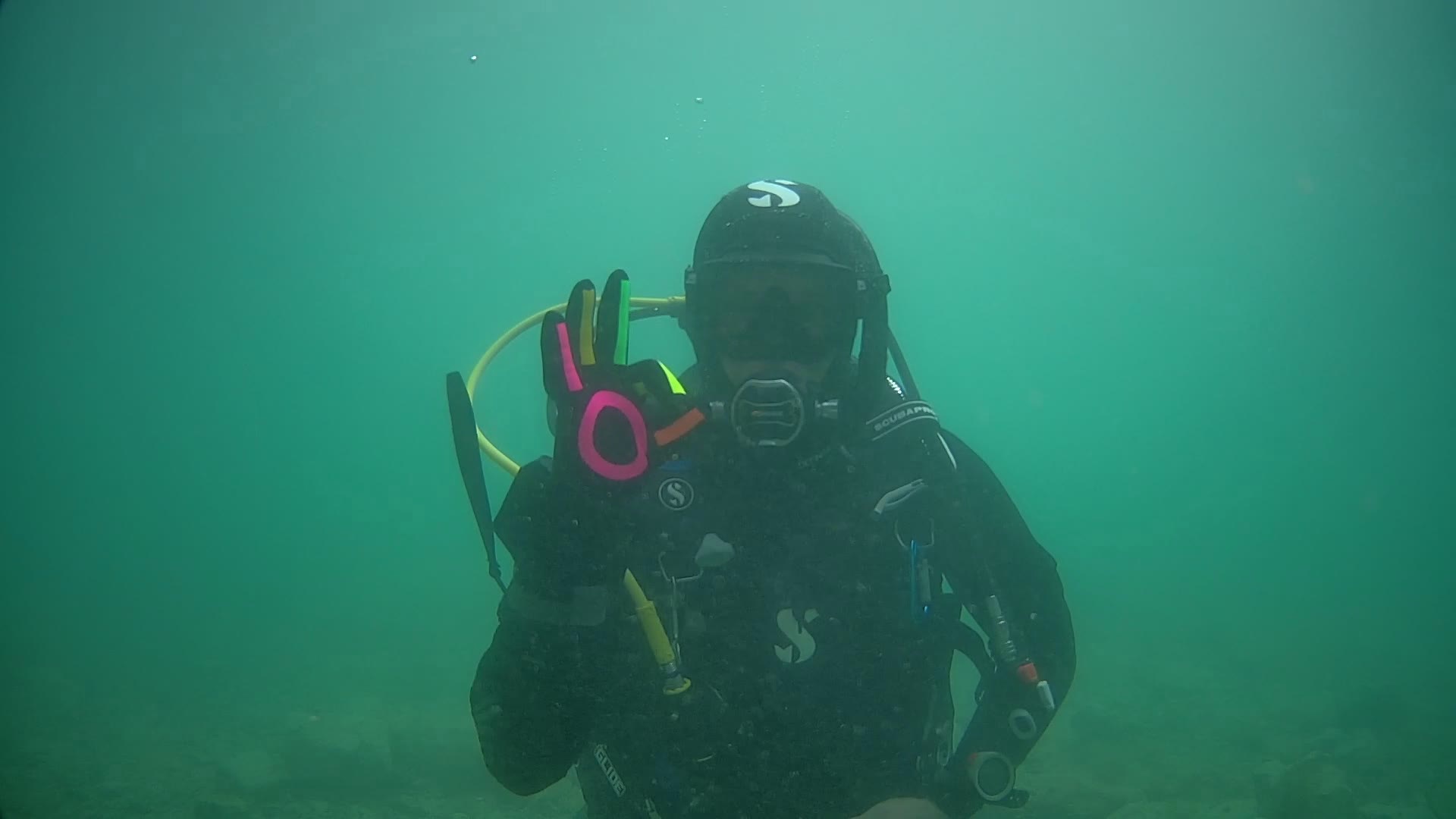}}\\
{\textbf{Reserve} \\ Semantics: air on reserve (50 bar left) \\ Type: static gesture \\ Hands: R \\ Palm/back: palm to camera \\ Fingers: fist} & {\includegraphics[width=0.49\textwidth]{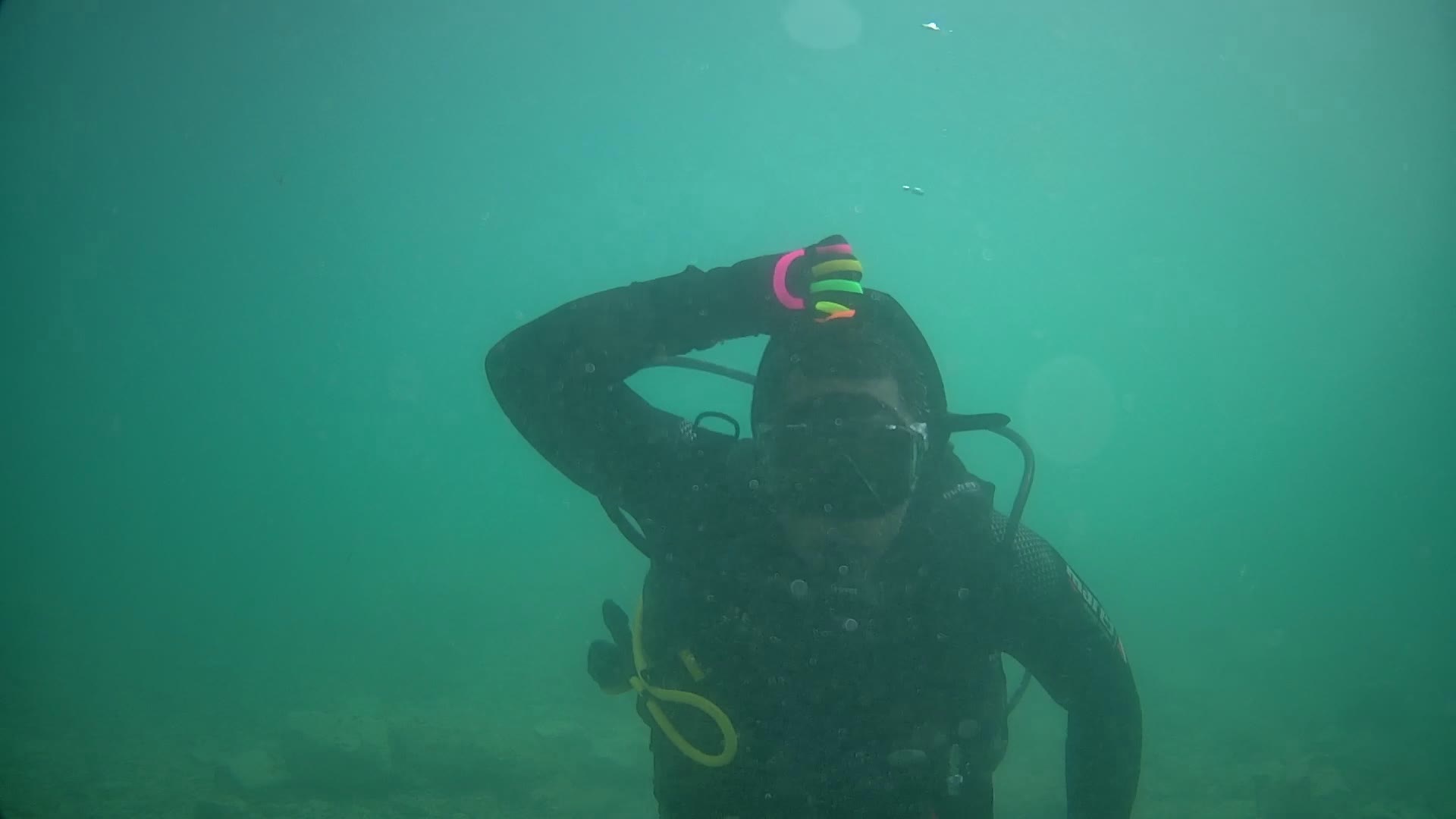}}\\
{\textbf{Turn} \\ Semantics: turn for 180° degrees \\ Type: dynamic gesture \\ Hands: R \\ Palm/back: back to camera \\ Fingers: 2R} & {\includegraphics[width=0.49\textwidth]{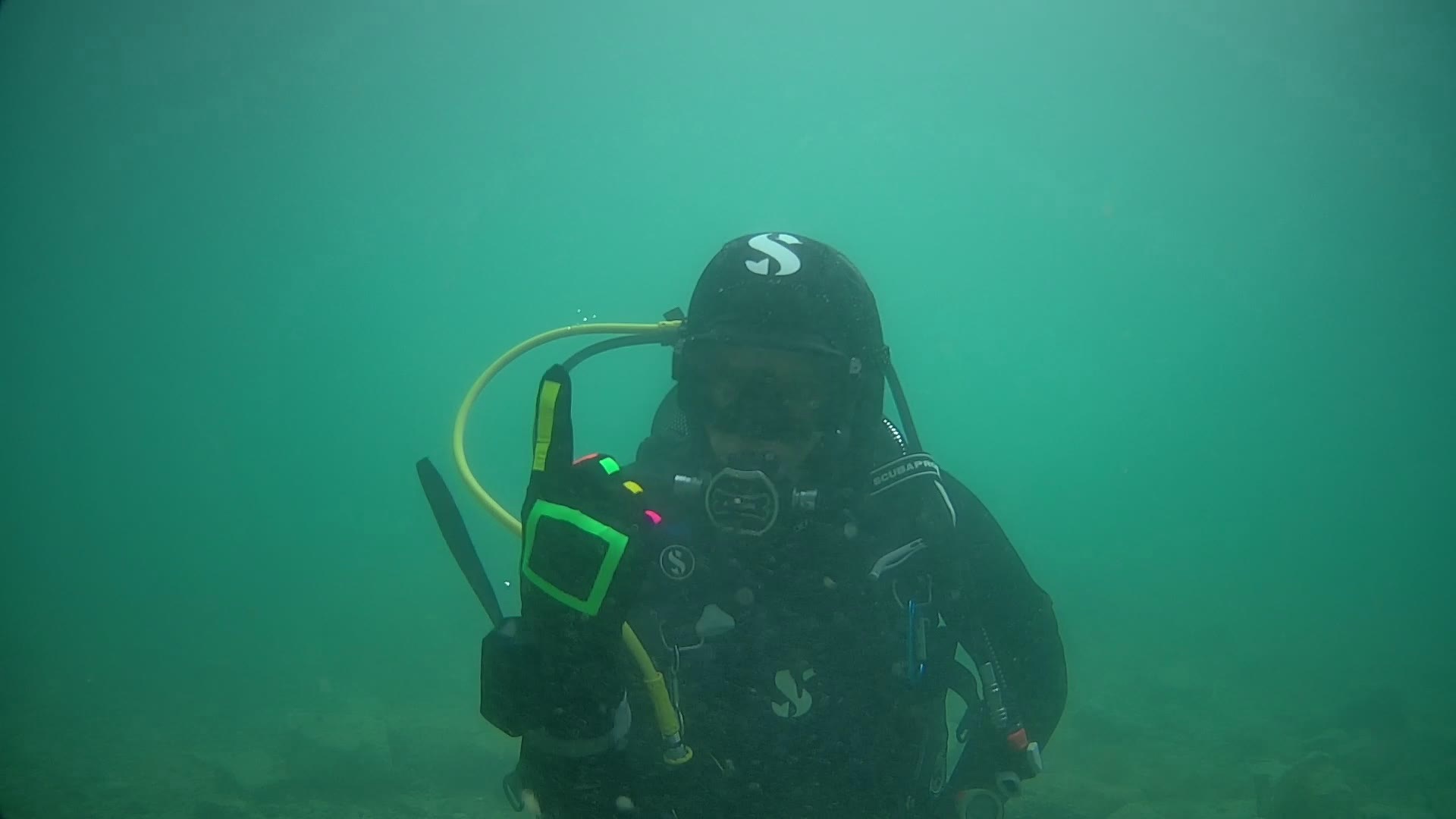}}\\
\end{longtblr}

Additionally, the main gesture that carries actual semantics sent in each message has accompanying gestures that include start of gesture recognition and start and end of the gesture. The first mentioned gesture starts the gesture recognition process on the smart diving glove and it is a gesture that cannot be easily done by accident during regular diving activities. The start and end of a message is the same "look at me" gesture, intended to "divert the robot's attention" to the diver and in a way safeguard the main message to be transmitted. Example of these gestures together with gesture meaning, semantics and instructions for performing the gesture can be found in Table \ref{geste_za_inicijalizaciju}.

\begin{table}[ht]
\caption{Communication safeguard gestures for starting and stopping the gesture recognition process and sending the acoustic command from the glove.}
\label{geste_za_inicijalizaciju}
\centering
\begin{tblr}{
  colspec = {X[c,h]X[c]},
  stretch = 0,
  rowsep = 2pt,
  hlines = {1pt},
  vlines = {1pt},
  colsep = {1pt},
}
\textbf{Gesture} & \textbf{Image}\\
{\textbf{Start communication} \\ Semantics: start gesture recognition \\ Type: static gesture \\ Hands: 2R, 5R \\ Palm/back: back to camera \\ Fingers: 1R} & {\includegraphics[width=0.49\textwidth]{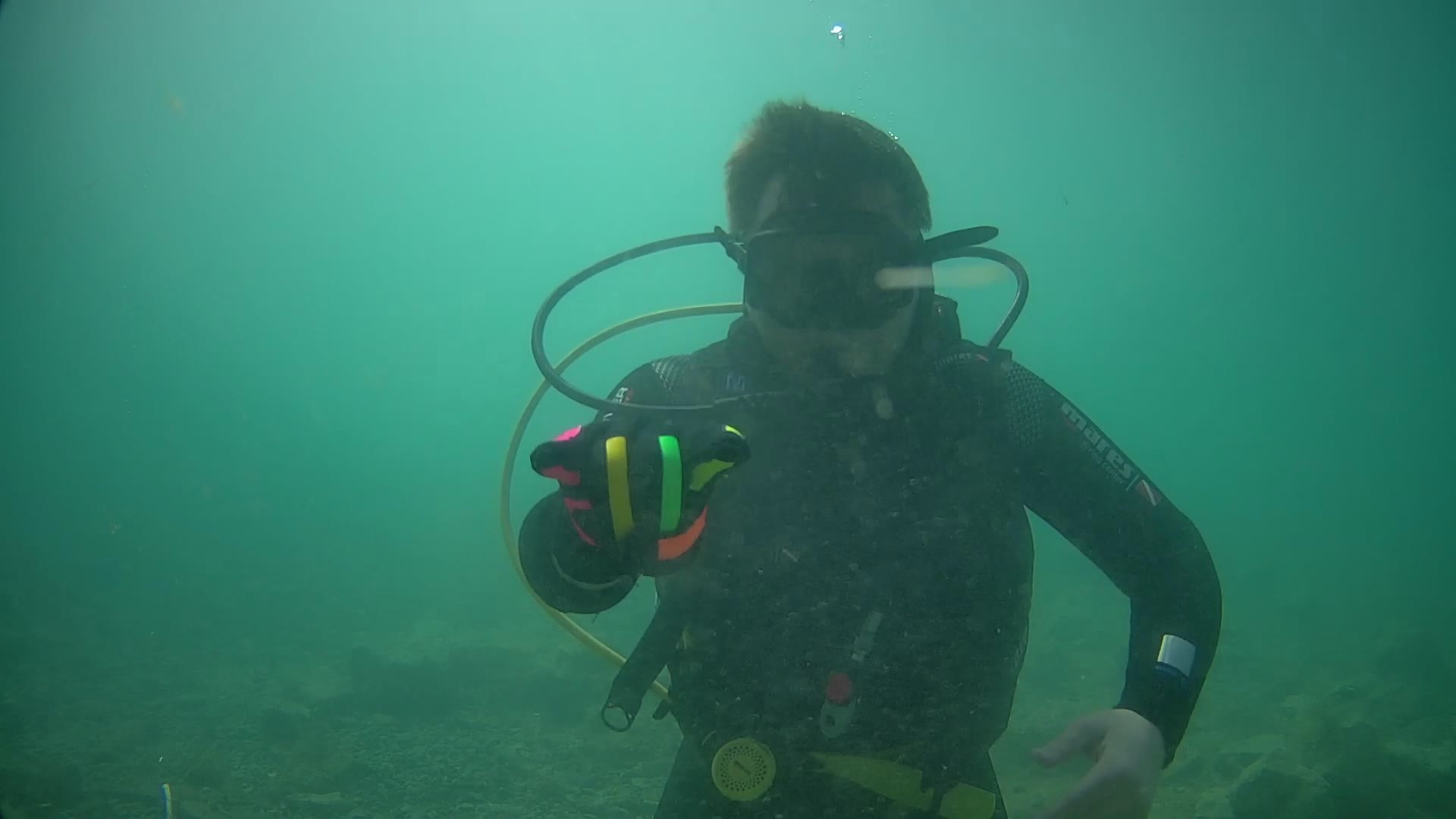}}\\
{\textbf{Look at me} \\ Semantics: start/end command \\ Type: static gesture \\ Hands: R \\ Palm/back: back slightly towards the camera \\ Fingers: 2R,3R} & {\includegraphics[width=0.49\textwidth]{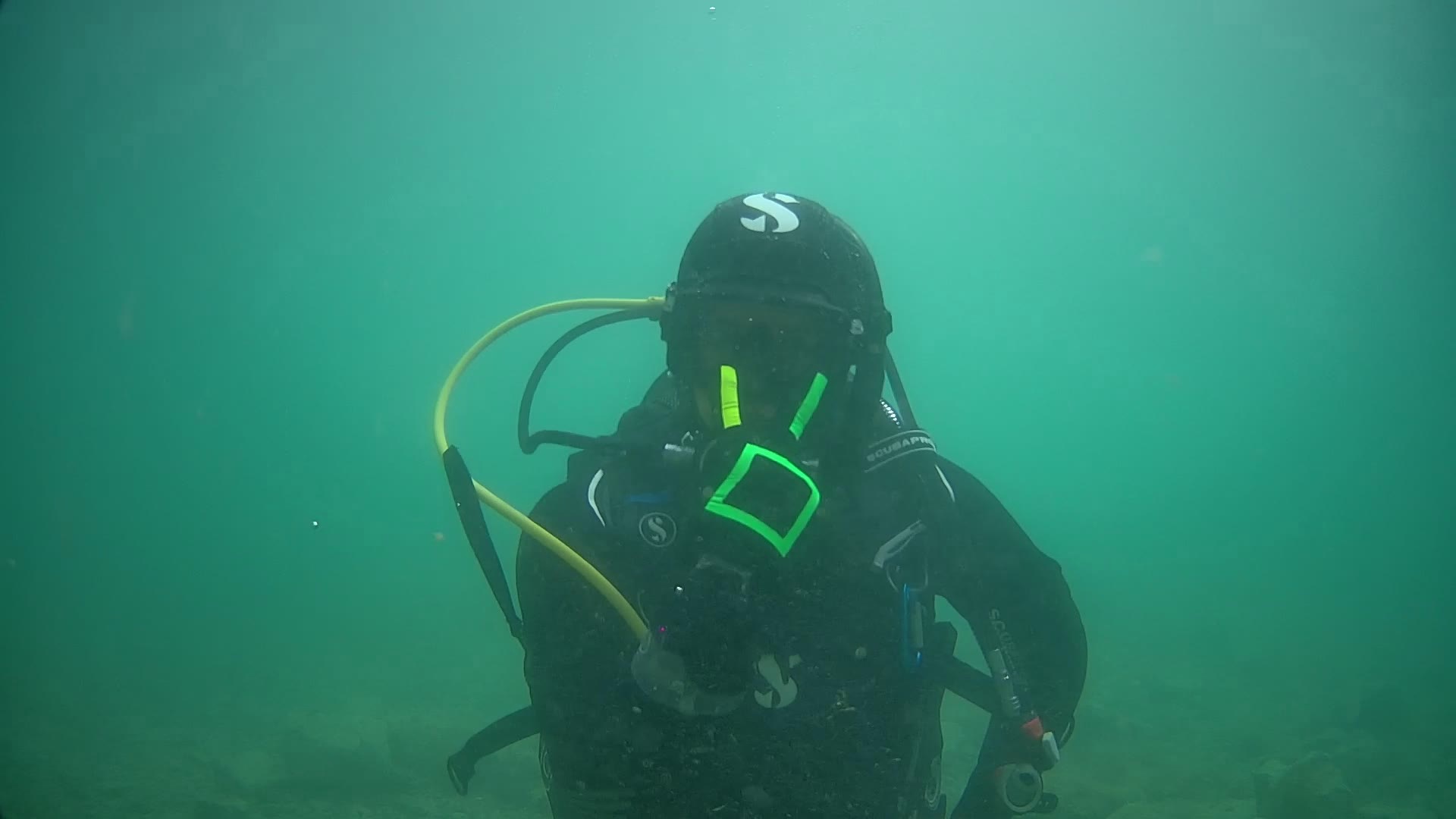}}\\
\end{tblr}
\end{table}

\subsection{Data collection} \label{datacollection}
\begin{figure}
\begin{center}
\includegraphics[width=0.7\textwidth]{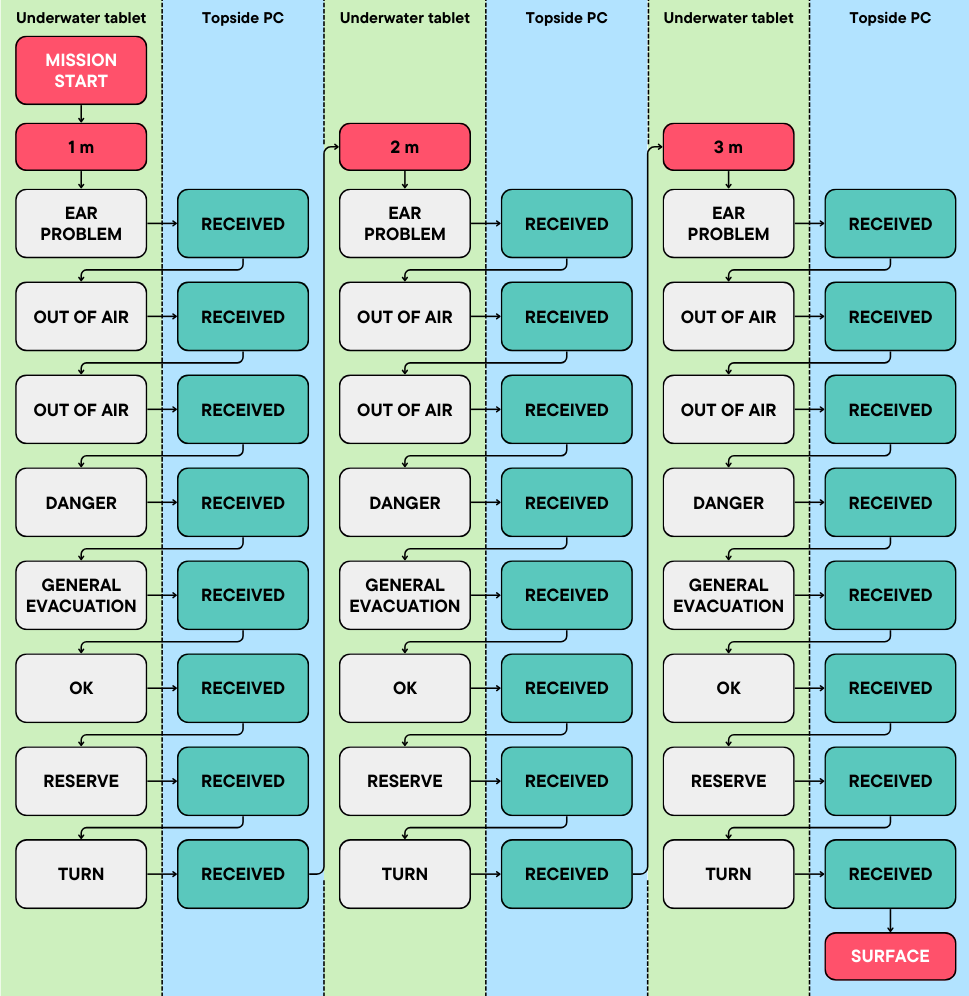} 
\caption{Workflow diagram of the underwater gesture experiment recording with tablet messages displayed to the diver shown in green columns, message receipts shown in blue columns. Each set of gestures is repeated at 1, 2 and 3 meters from the camera.} 
\label{fig:workflow}
\end{center}
\end{figure}

The data was collected in two separate locations during September and October 2022. In order to ensure diversity of the dataset and capture different visibility conditions, both pool and sea water conditions were used in the experiment. For the location of the sea part of the dataset Biograd na Moru in Croatia was chosen. The pool part of the dataset was recorded at the Laboratory for Underwater Systems and Technologies at the University of Zagreb, Croatia. For each of the two locations five different divers were recorded, both male and female. Before the recording mission began, the divers were given an introductory briefing describing the gesture recognition glove, the background of the project and the purpose of the data collection. Each of the gesture instructions were first demonstrated on screen and in a hands-on demonstration, as well as instructions on how to use the underwater tablet. Each participant first attempted to mimic all the gestures in a dry environment to ensure there was no doubt before going on the dive. Underwater, the divers were simply reminded of the sequence of gestures and the positioning distance, using pictures as a reminder of how to perform each gesture. The sequence of eight chosen gestures was repeated at all three distances from the camera with corresponding messages and instructions shown on the underwater tablet, as depicted in the workflow diagram on Figure \ref{fig:workflow}.

At the mission start, after checking all sensors and communications, data acquisition was started and the gesture instructions sequence starts displaying on the underwater tablet. The poolside PC records tablet instructions as a starting reference timestamp, together with camera video recording and data received from acoustic modems. Raw data from the sensors in the glove is stored directly on the glove and is downloaded using Bluetooth and synchronised at each mission end. For synchronising video with the rest of the dataset, a microsecond timer was recorded from the acquisition PC at multiple time difference samplings of each mission and averaged. The video delay could be identified subtracting the timestamp of a particular frame and the time recorded on the frame image. The data received acoustically from the diving glove is synchronised to the exact time at which the gesture was made $Tg$ using Expression \eqref{delay}.

\begin{equation}
Ta=Tg+0.330s+3/1520s\label{delay}
\end{equation}

When an acoustic message arrives at the receiver, only the arrival timestamp $Ta$ is known and the transmission delay has to be subtracted. The communication protocol of the acoustic modem defined in \cite{nanomodem} uses a data structure composed of a 30 ms synchronisation chirp, followed by a 75 ms message header, data frame which for our 2 bytes of transmitted data lasts 25 ms, and finally 16 parity bytes lasting 200 ms. The last factor is to account for travel time through water, where the receiving modem was placed laterally roughly 3 m away from the diver position and 1520 m/s that was taken as the speed of sound through water. As that factor was close to negligible ($\sim$2ms) and it is likely that regardless of that correction it would fall under the same video frame so no further adjustments were necessary there.

\subsection{Data Extraction and Processing}
After synchronising all the sensors, the timing data was merged in a single .csv file using a Python script for processing. The script merges tablet messages, acoustic messages, and glove gesture detection data with the list and timestamps of all video frames in a large dataframe. Once assembled, it was possible to detect and determine exact times when particular gestures occurred and find corresponding video frames. For each detected gesture data the script decodes tablet, acoustic and glove data into their semantic meanings. The rest of the dataframe, i.e. the rows that did not contain actual gesture data on any of the sensors where discarded for better readability. This filtered dataframe was used as the base for extracting video frames for the final gesture dataset. For a gesture detected on any of the sensors, the closest video frame was found using timestamps as a key. 
Since some of the gestures consist of dynamic movement, a snippet of many video frames was extracted for each detected gesture. Reviewing footage of such gestures showed that 3 seconds cover all the integral movements needed to form an individual gesture in most of the cases. As we already determined the exact time at which a gesture detection happened on the glove, we included all the frames in the duration of 3 seconds around that time and in the following ratio; 3/4 of the frames that preceded the moment of detection and 1/4 of frames that followed after the detection happened on the gesture recognition glove. That means that each snippet contains, depending on the frame rate, all video frames starting from 2.25 seconds before a gesture was detected up to 0.75 seconds after the detection happened. This methodology proved to cover the essential movements needed to compose any particular gesture used in this experiment.

\subsection{Database structure}
\begin{figure}
\begin{center}
\includegraphics[width=0.7\textwidth]{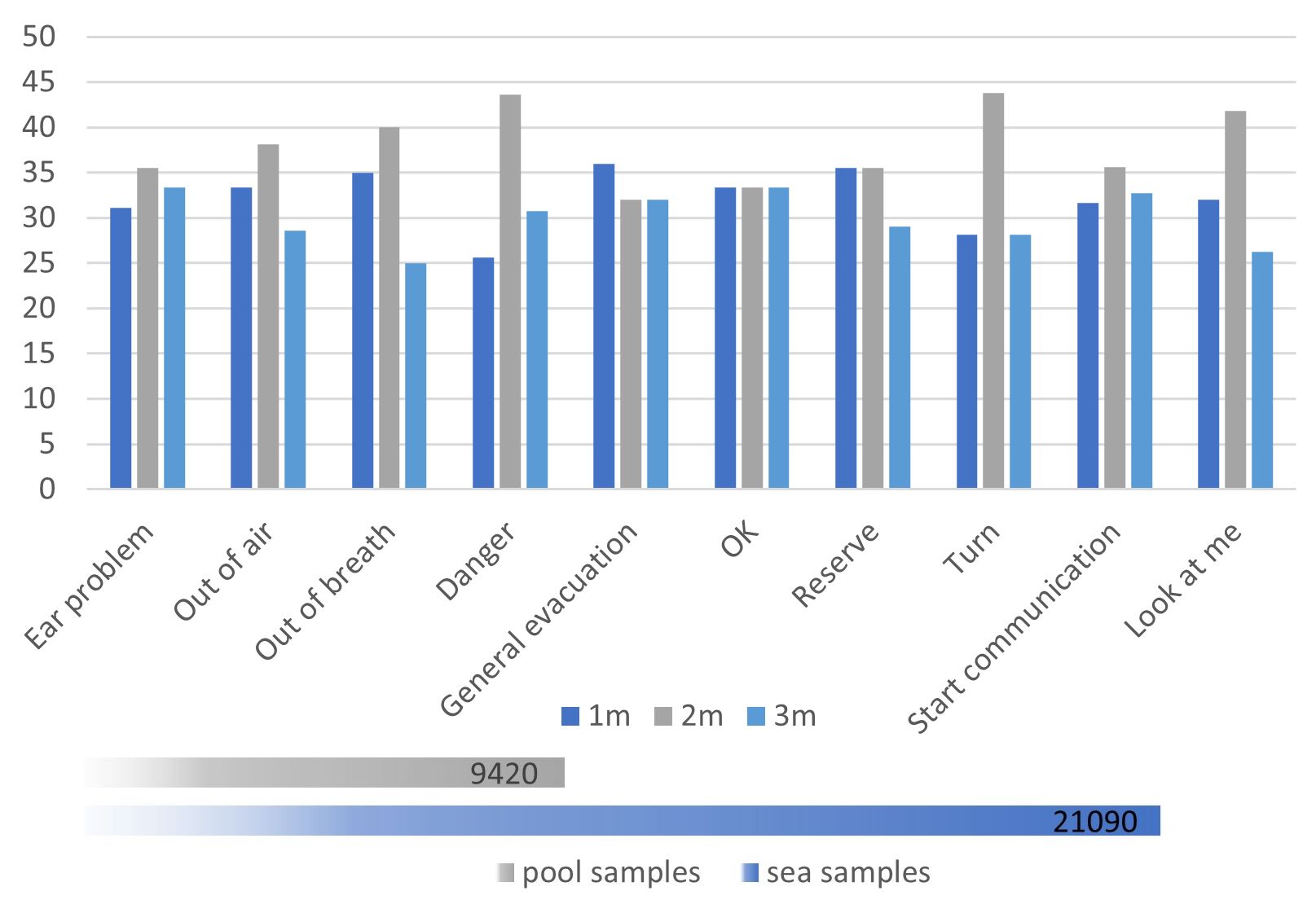} 
\caption{Relative distribution of image samples per gesture by distance from the camera (above) and total image samples per environment (below).} 
\label{fig:samples_distribution}
\end{center}
\end{figure}

The dataset presented in this paper is structured into folders, each representing a distinct gesture. Each gesture folder consists of numbered subfolders containing instances of that particular gesture, comprising of snippet frames that encapsulate temporal aspects of the gesture's execution. This systematic organization is intentionally designed to optimize compatibility with machine learning and visual detection techniques.\textcolor{black}{ Supplementary to the dataset organised in folders, we have supplied a metadata .csv file linking each gesture instance to relevant collection attributes such as environment (sea or pool), diver distance from the camera (1m, 2m or 3m), diver identifier and number of samples in instance snippet.} The hierarchical arrangement ensures a clear and efficient framework for these techniques to navigate through the dataset. By adopting such a structure, researchers and practitioners can harness the full potential of the dataset for training and evaluating gesture recognition models, as well as for refining visual detection algorithms. This approach aims to streamline accessibility and enhance the dataset's utility in the context of advanced computational analyses so they can be compared to the gesture detection onboard the smart diving glove used in this experiment. Exact numbers of successfully recorded frames of each particular gesture are given in Table \ref{broj_gesti} \textcolor{black}{ and distribution of samples per distance and environment on Figure \ref{fig:samples_distribution}}.

\begin{table}[ht]
\caption{Distribution and number of successfully detected images per gesture}\label{broj_gesti}
\resizebox{\textwidth}{!}{
\begin{tabular}{ccc}
\hline
\textbf{Gesture} & \textbf{Number of gestures in dataset} & \textbf{Number of frames}\\
\hline
Ear problem & 37 & 1350 \\
Out of air & 18 & 630\\
Out of breath & 17 & 600\\
Danger & 35 & 1170\\
General evacuation & 24 & 750\\
OK & 18 & 630\\
Reserve & 28 & 930\\
Turn & 28 & 960\\
Start communication & 250 & 8420\\
Look at me & 438 & 15000\\
\hline
\textbf{TOTAL} & \textbf{892} & \textbf{30440}\\
\hline
\end{tabular}}
\end{table}

\subsection{Data description}

\begin{figure}[H]
\begin{center}
\includegraphics[width=\textwidth]{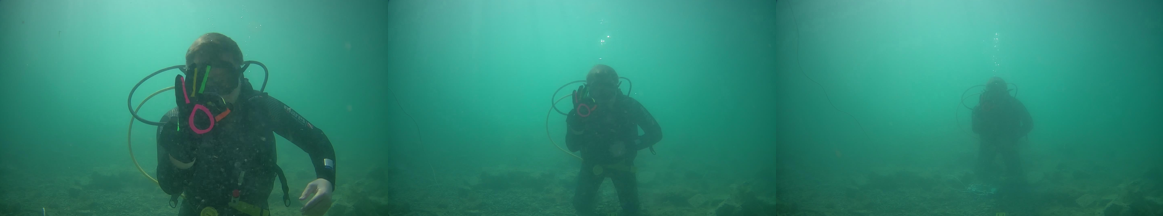} 
\caption{"OK" gesture at 1, 2 and 3 meters from the camera in the sea.} 
\label{fig:screenshotsea}
\end{center}
\end{figure}

\begin{figure}[H]
\begin{center}
\includegraphics[width=\textwidth]{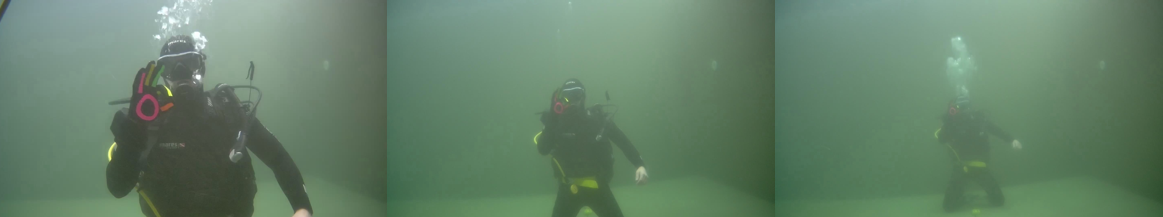} 
\caption{"OK" gesture at 1, 2 and 3 meters from the camera in the pool.} 
\label{fig:screenshotpool}
\end{center}
\end{figure}

The dataset encompasses a diverse collection of image snippets, each lasting for a duration of 3 seconds, strategically recorded at distances of 1, 2, and 3 meters from the camera. This variation in proximity ensures a comprehensive representation of gestures captured in different spatial contexts and visibility conditions. The image snippets are captured to include both pool and sea environments, thereby introducing real-world complexities into the dataset. The images are in the camera's native resolution of 1920x1080 pixels and are in no way pre-processed in terms of brightness, color correction, image distortion etc. Examples of images of a diver performing the "OK" gestures at 1, 2 and 3 meters from the camera can in the sea and in the pool be seen on Figures \ref{fig:screenshotsea} and \ref{fig:screenshotpool}, respectively. This standardized resolution aims at streamlining usage of the dataset for machine learning purposes, enabling the development and evaluation of robust gesture recognition models. The deliberate consideration of various factors in the dataset design aims to enhance its applicability in real-world scenarios and contribute to the advancement of underwater gesture recognition in the context of diver-robot or diver-diver interaction.
 

\subsection{Smart glove recognition statistics} \label{recognition_statistics}
As described in subsection \ref{datacollection}, in parallel to recording the image and video of the divers performing gestures, all the hand movements are recorded using the smart diving glove with sensors incorporated in each of the fingers and back of the palm and the detection algorithm described in \cite{remote_experiments}. In order for the algorithm to identify an intentional gesture, a particular sequence of gesture has to be performed. This consists of a "start gesture recognition" gesture, followed by a "look at me"  gesture after which comes the main gesture that carries actual message semantics. Once the main gesture is successfully detected, the diver repeats a "look at me" gesture which ends the detection process and transmit the gesture command acoustically. An entire gesture sequence example for the "ear problem" message is shown in Table \ref{gesture_sequence}.

\begin{table}[H]
\caption{Semantics of a complete sequence required to detect a gesture on the recognition glove on the "ear problem" gesture example, with gesture description in the top row and gesture meaning in the bottom row.}\label{gesture_sequence}
\begin{tblr}{
  colspec = {X[c,h]X[c]X[c]X[c]},
  stretch = 0pt,
  rowsep = 2pt,
  hlines = {1pt},
  vlines = {1pt},
  colsep = 1pt,
}
\includegraphics[width=0.24\textwidth]{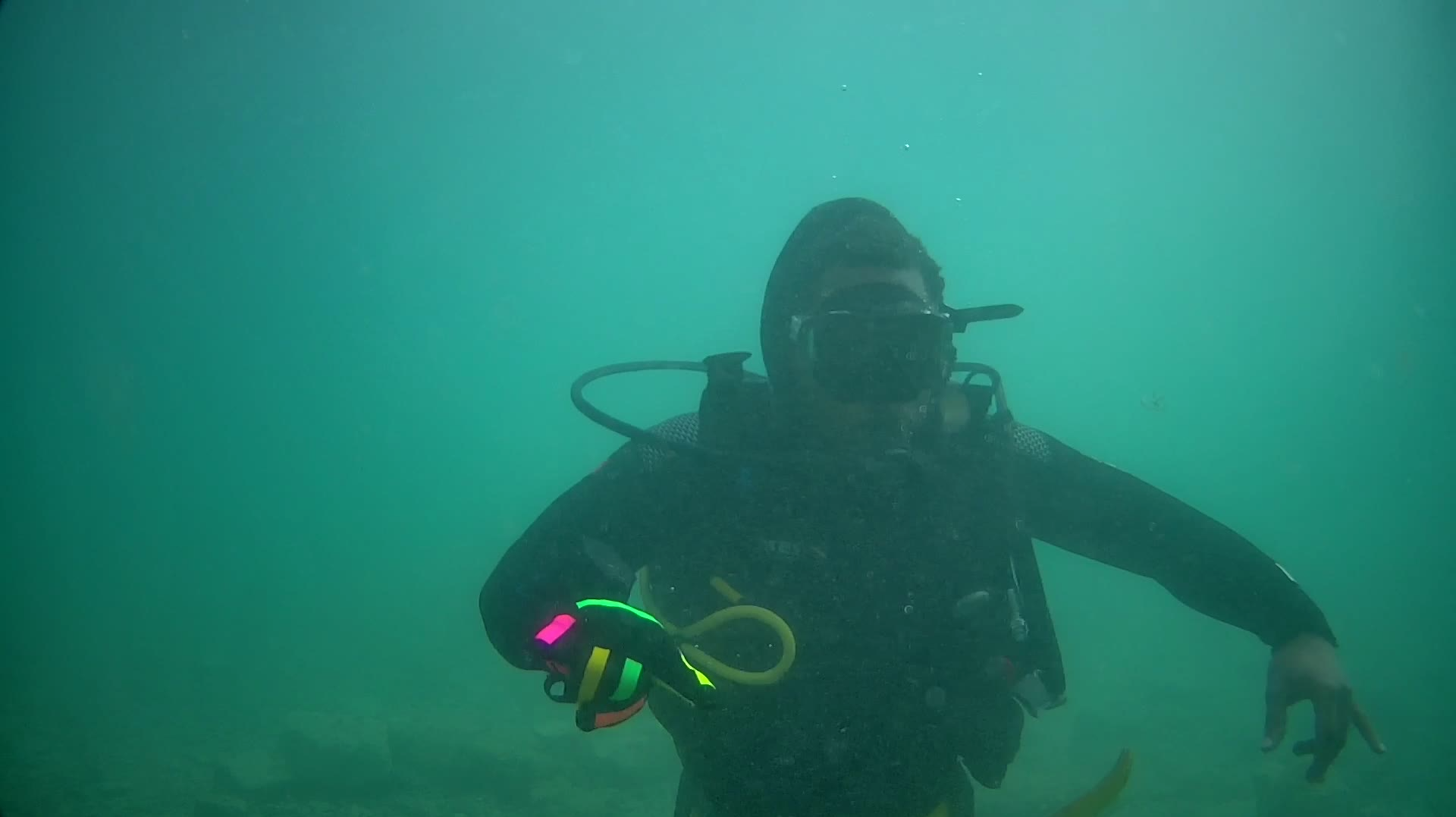} &  \includegraphics[width=0.24\textwidth]{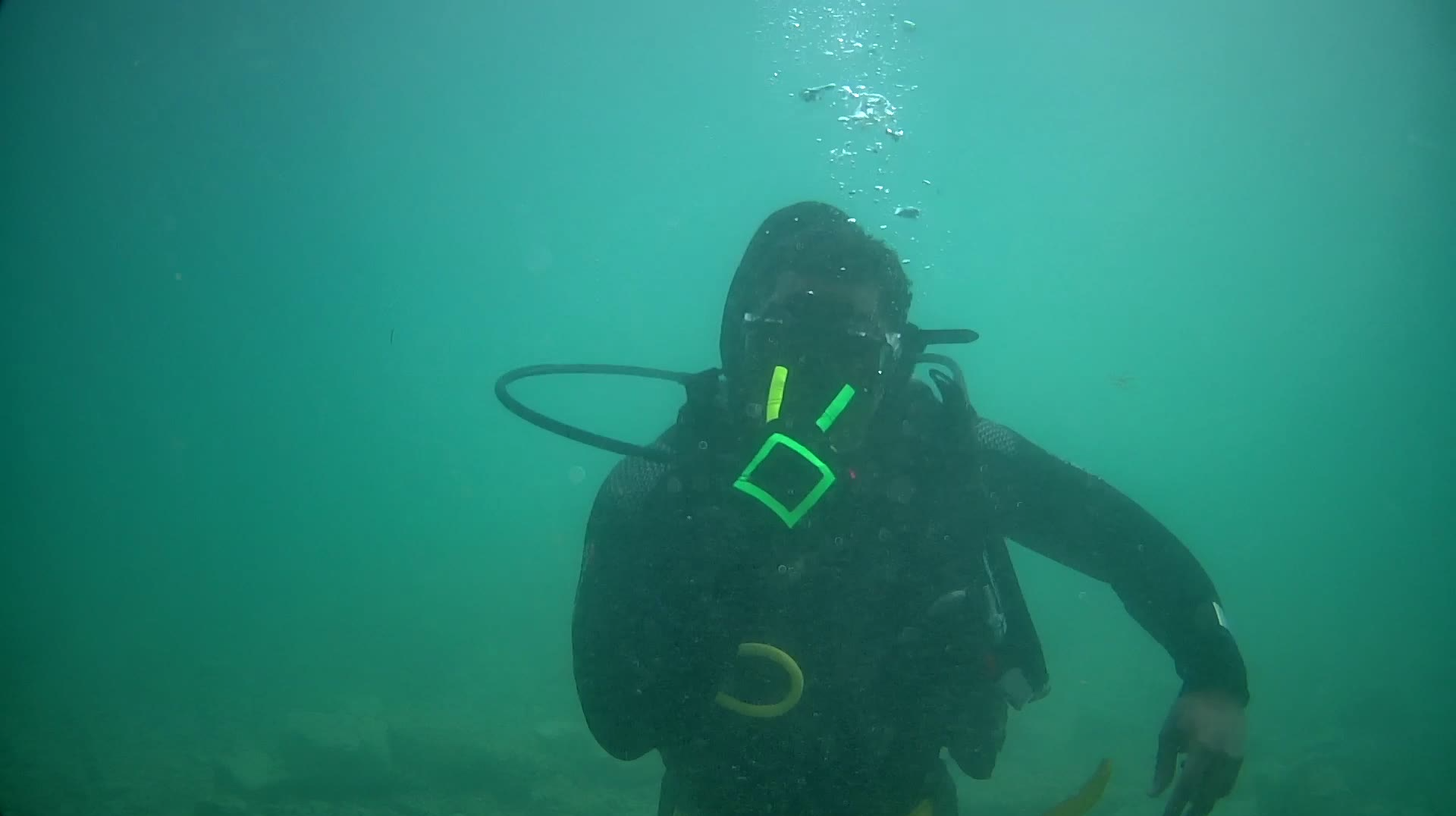} &\includegraphics[width=0.24\textwidth]{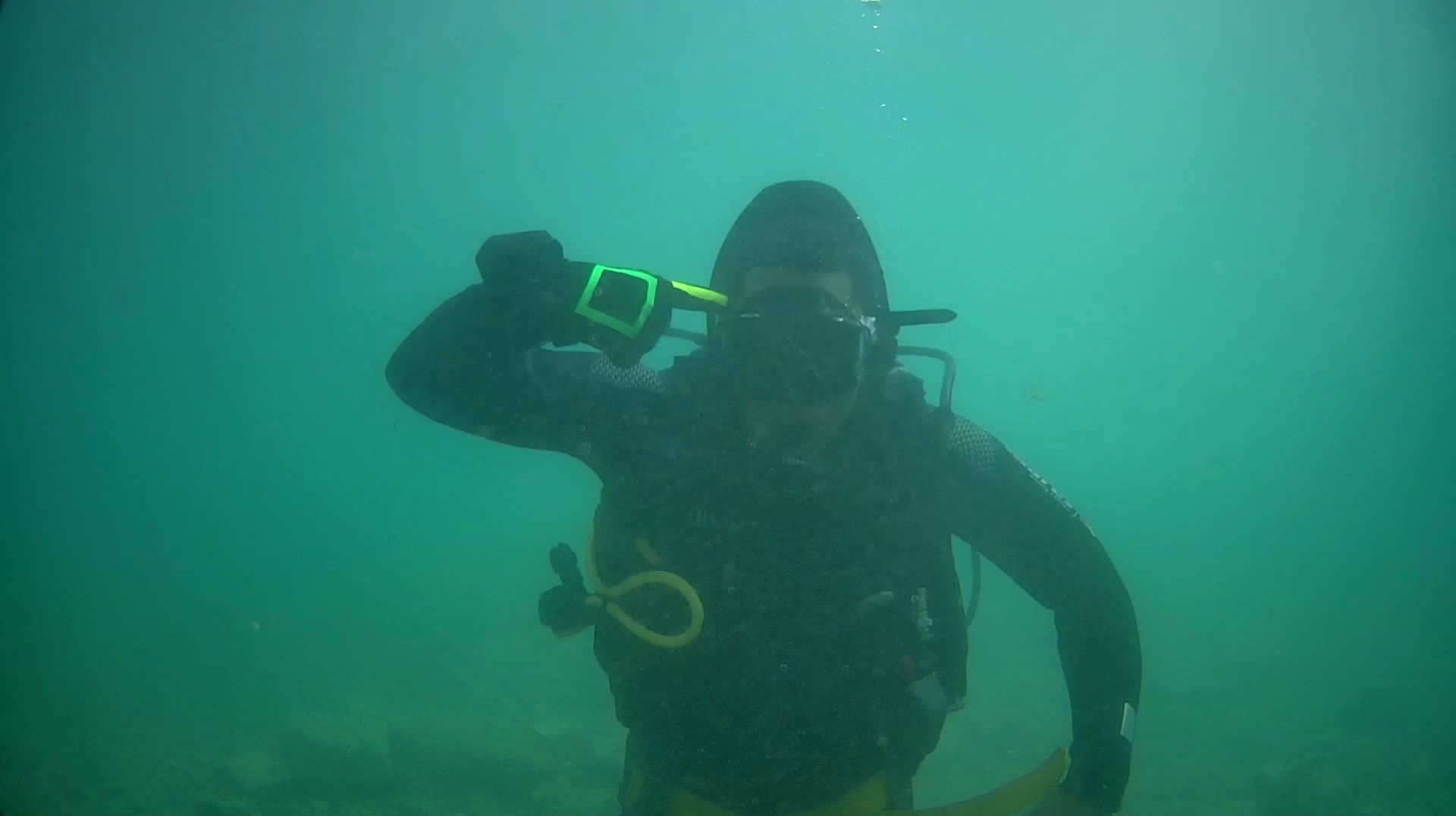} &  \includegraphics[width=0.24\textwidth]{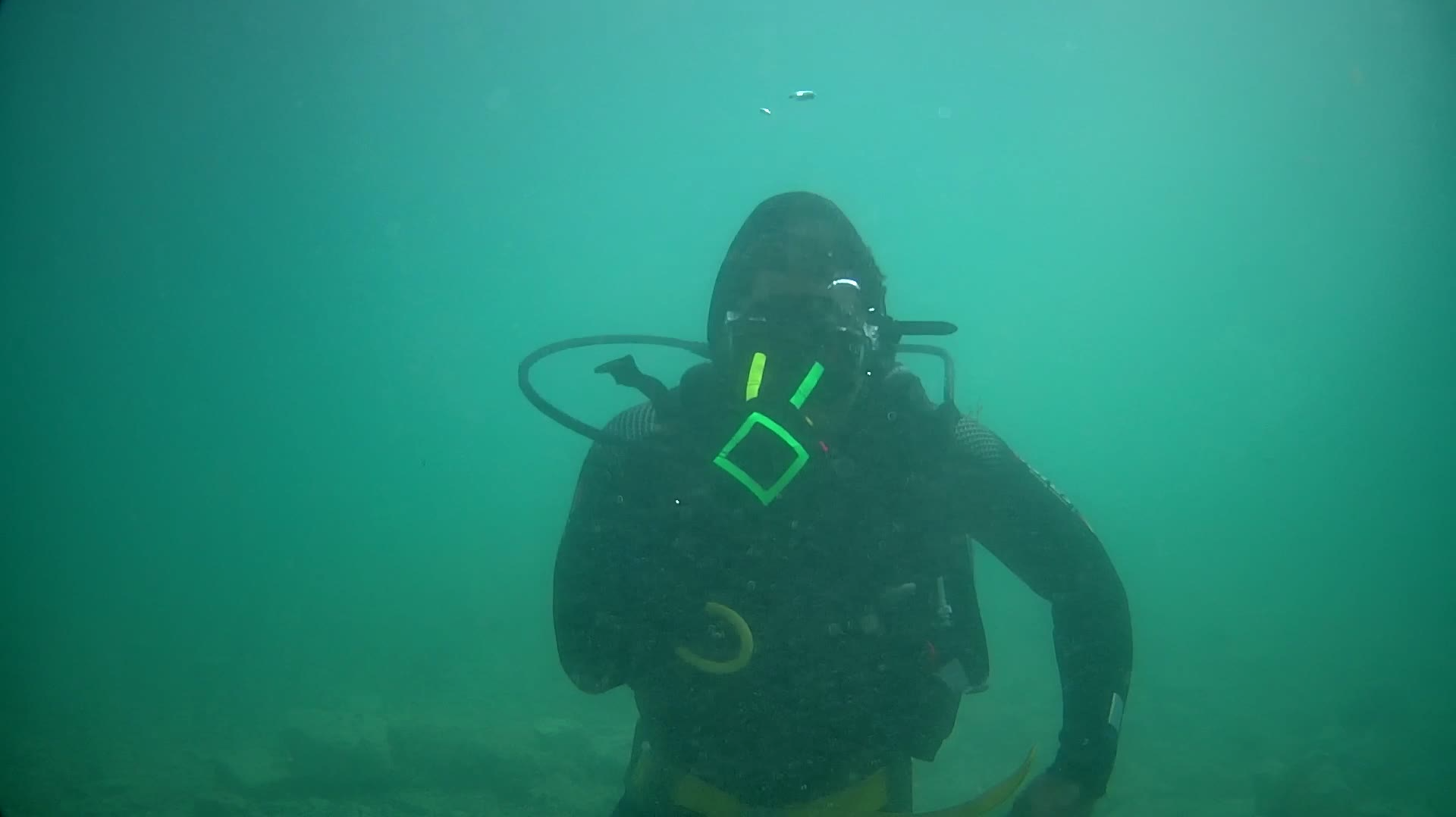}\\
"Rockon" gesture & Look at me &  Ear problem & Look at me\\
Start gesture recognition & Start command & Semantics carrying gesture & End command \\
\end{tblr}
\end{table}

The added overhead, in addition to the modulation process and acoustic time of flight, adds significantly to the time needed to transmit each gesture compared to a purely visual detection method. Table \ref{gesture_durations} presents average time needed to complete a gesture for each of the participants of the experiment. The second column analyzes the reaction time which is the time elapsed between issuing the next gesture instruction on the underwater tablet and the time when the diver starts performing that gesture. "Gesture time" refers to the time needed for the diver to complete the entire gesture sequence, including the "start gesture recognition", "look at me", main gesture and the final "look at me" gesture that ends recording. Timing on the glove starts when the first gesture (start gesture recognition) is recognized, which also initiates the sequence to store sensor values on the glove, and finishes when the last gesture is recognised, triggering command transmission. Transmission times account for the time from when the gesture is detected and sent from the glove to the moment when it is successfully received at the poolside PC. The last column shows the average total time, from giving the tablet instruction visual to receiving the detected gesture on the acoustic modem.

\begin{table}[ht]
\caption{Average gesture duration across all the users and overall in seconds}\label{gesture_durations}
\begin{center}
\resizebox{\textwidth}{!}{
\begin{tabular}{ccccc}
\hline
\textbf{User} & \textbf{Reaction time} & \textbf{Gesture time} & \textbf{Transmission time}  & \textbf{Totoal gesture time}\\
\hline
1 & 3,117 & 10,079 & 0,328 & 13,524 \\
2 & 9,337 & 11,218 & 0,326 & 20,880\\
3 & 4,396 & 9,665 & 0,327 & 14,389\\
4 & 5,637 & 8,000 & 0,333 & 13,971\\
5 & 3,547 & 7,919 & 0,335 & 11,801\\
6 & 5,622 & 10,512 & 0,330 & 16,465\\
7 & 3,288 & 6,252 & 0,331 & 9,872\\
8 & 3,582 & 10,062 & 0,321 & 13,965\\
9 & 3,190 & 6,687 & 0,329 & 10,206\\
10 & 4,992 & 16,804 & 0,331 & 22,127\\
\hline
\textbf{Average} & \textbf{4,671} & \textbf{9,720} & \textbf{0,329} & \textbf{14,720}\\
\hline
\end{tabular}}
\end{center}
\end{table}

\begin{figure}[H]
\begin{center}
\includegraphics[width=0.5\textwidth]{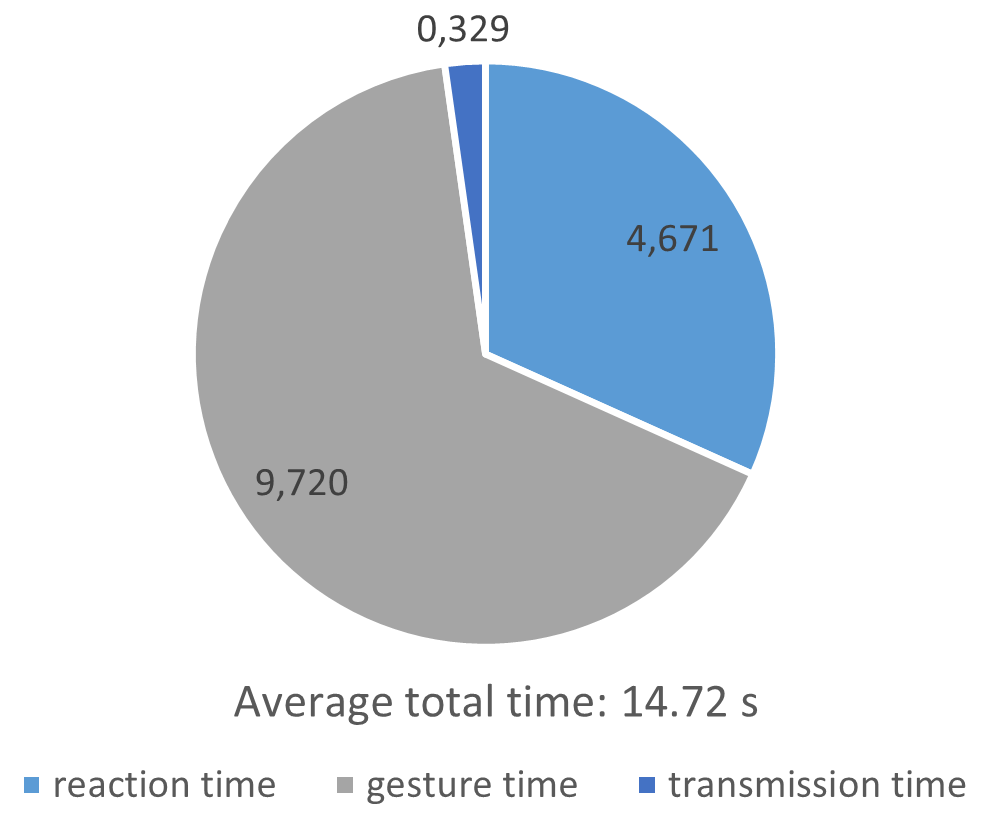} 
\caption{Time needed to complete and transmit a gesture command in seconds split in their respective contributions} 
\label{fig:gesture_time}
\end{center}
\end{figure}

In Figure \ref{fig:gesture_time} we can see more clearly the contribution of the individual components to the total gesture time. While the majority of the time amounts to performing the gesture sequence, if we do not consider reaction time, which should not depend on the communication modality, the diver needs on average 10 seconds to communicate a command to the underwater robot, another diver or a topside operator. While this may sound like a considerable amount of time, this form of communication has the benefit of not depending on visibility conditions, direct line of sight and it functions across appreciably larger distances than any visual method. Acoustic transmission time is not a significant factor, at least not at shorter distances as in this experiment.

\begin{figure}[H]
\begin{center}
\includegraphics[width=\textwidth]{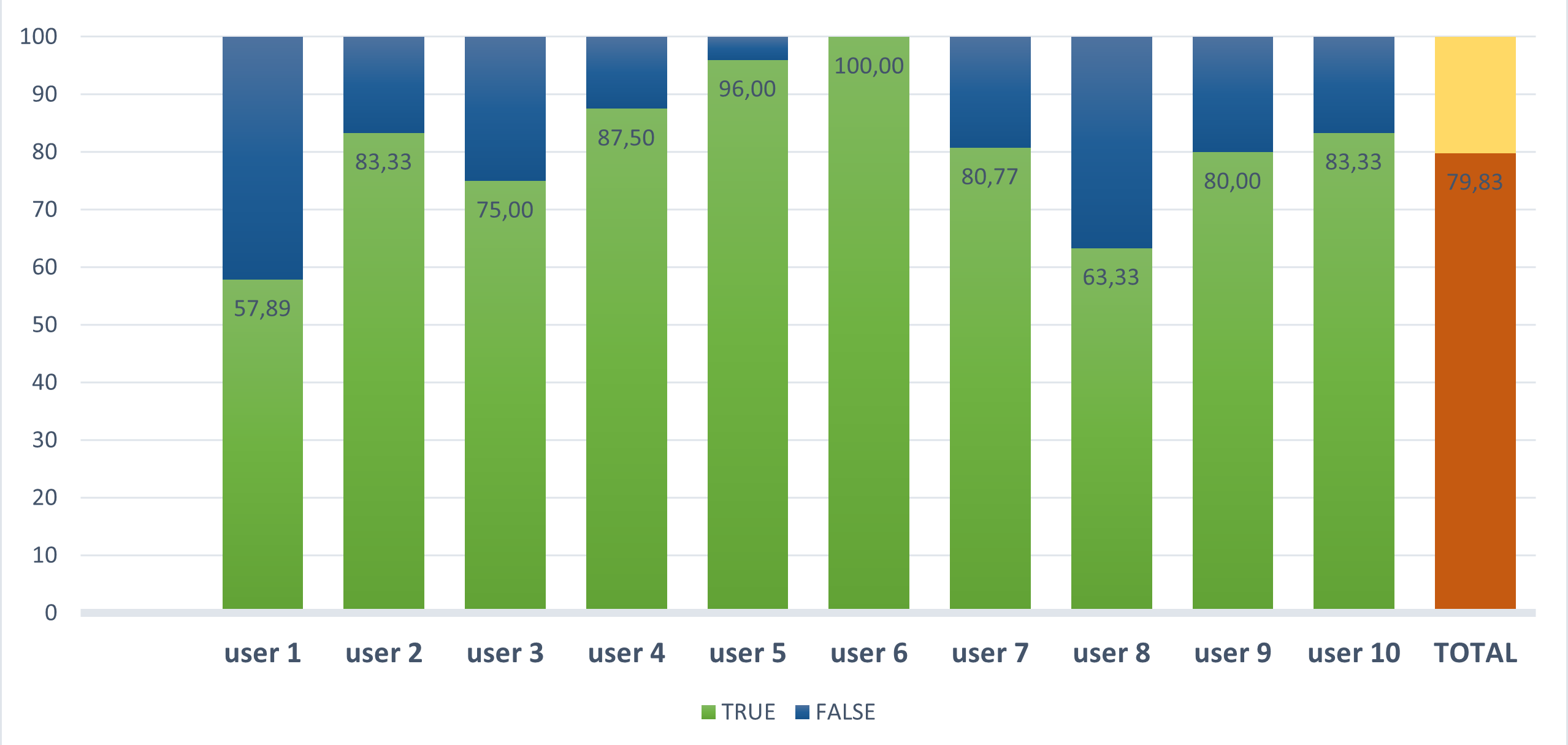} 
\caption{Percentage of gestures successfully detected and received through acoustics across users and overall} 
\label{fig:succesfully_received}
\end{center}
\end{figure}
\begin{figure}[H]
\begin{center}
\includegraphics[width=\textwidth]{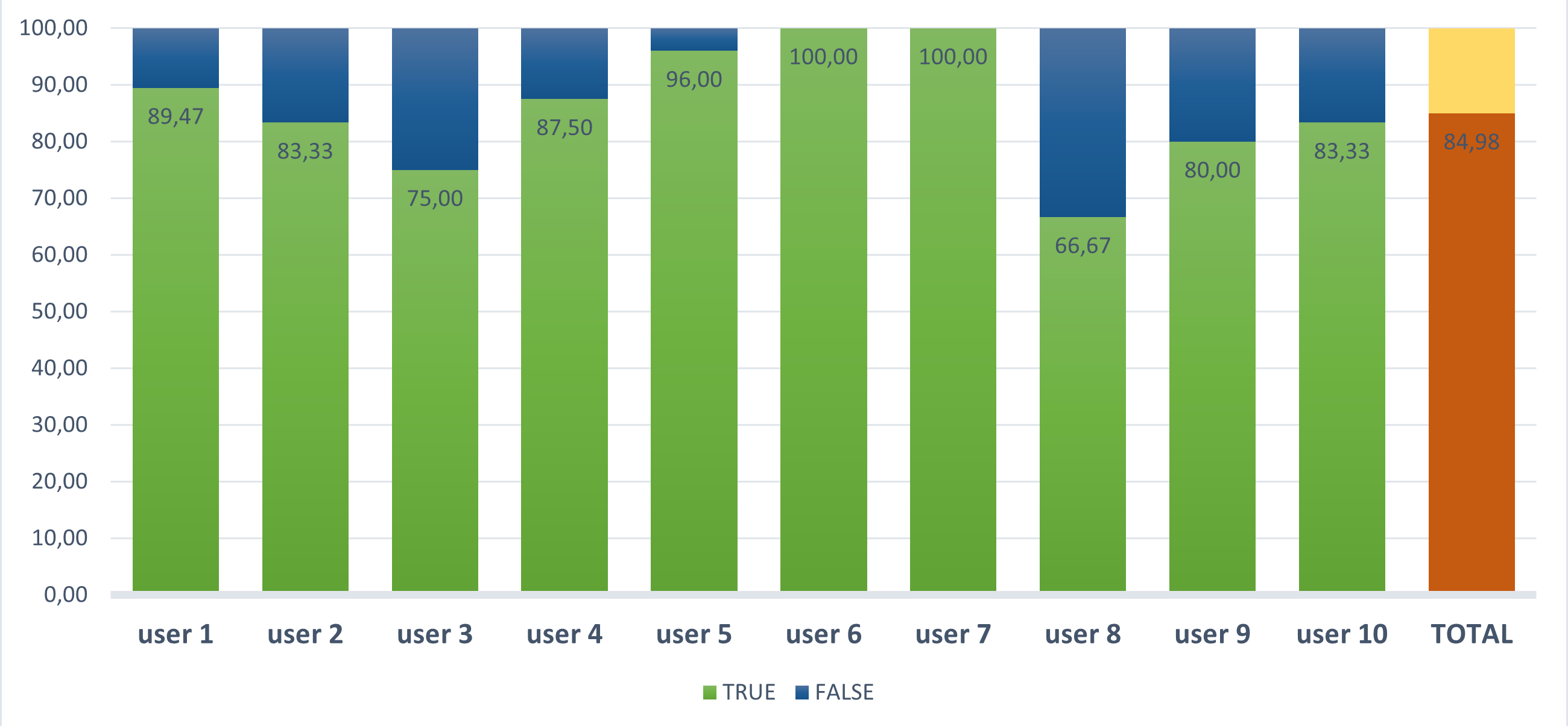} 
\caption{Percentage of gestures successfully detected on the gesture recognition glove across users and overall} 
\label{fig:sucesfully_detected}
\end{center}
\end{figure}

The gesture recognition using the smart diving glove and acoustic transmission does, however, come with it's own set of limitations. Since the participants of the experiment were not professional divers and had only a short briefing and instructions on how to use the glove, they did not have extensive practice in performing the gestures. As a result, not all gestures were done correctly and some of them were not detected. As is to be expected, the performance varied significantly between participants. The percentage of gestures that were detected and received on the poolside PC can be seen on Figure \ref{fig:succesfully_received}. In addition to the unsuccessful detections on the glove itself, some of the gestures that were detected on the glove and their message was sent, were not received at the acoustic modem. Since the experiments were conducted in the pool and in the sea both near a flat wall, there was a possibility for the acoustic signal to suffer from multipath issues. The situation is even worse in the pool where all the surrounding surfaces were flat, which is less than ideal for underwater acoustics \cite{multipath}. To visualize the number of gestures that were detected successfully but failed to transmit, Figure \ref{fig:sucesfully_detected} contains total percentages of successful detection on the glove, including ones that were not received at the PC end. The last column of both figures shows the overall average of successful detections. We can see that on average nearly 85\% of the gestures are detected successfully between all users but only 79.83\% of them are received at the acoustic modem as gesture commands.

\section{Conclusion}

The aim of the research described in this article was to create a comprehensive dataset of diving gestures with the purpose of benchmarking different gesture recognition algorithms. For the gesture base, the CADDIAN language was used, a language based on consolidated and standardized underwater gestures commonly used in recreational and professional diving and published by Chiarella et. al. in \cite{caddian2}. The dataset presented there and the supporting papers describe in detail the list of gestures and their meaning with instructions on how to reproduce and interpret them. Since CADDIAN gestures were intended to be visually processed by underwater robots, colored strips were added to the standard diving glove to enhance the visual recognition. During our previous research projects, a capacitance-based sensor glove was developed with the capability of detecting gestures in real time and transmitting them via acoustic communication. To provide basis for comparison of this gesture recognition method with visual recognition, such a glove, retrofitted with corresponding colored strips, was used and programmed to recognise selected gestures form the CADDIAN language. Since the processing on the glove itself has limited computing power and the transmission via acoustics entails a delay, the creation of a comprehensive dataset would be a good basis for the comparison between different modalities of machine gesture recognition. In total, five divers performed the gestures in seawater and five in freshwater, at three different distances from the camera. The result is over 30,000 images with almost 900 gestures recorded both in images and in the sensors of the smart diving glove. The statistics of the recognition success on board the smart diving glove were presented and analyzed. This will hopefully provide a solid basis for comparing different recognition techniques and quantifying their performance in terms of speed, accuracy and success rate.

\section*{CRediT authorship contribution statement}{Conceptualization, Đ.N. and C.W.; methodology, Đ.N., C.W., D.O.A. and I.K.; software, Đ.N. and D.O.A.; validation, C.W. and I.K.; formal analysis, I.K. and D.O.A.; investigation C.W., D.O.A. and I.K.; resources, Đ.N.; data curation, D.O.A. and I.K.; writing---original draft preparation, I.K.; visualization, I.K.; supervision, N.M. and I.A.; funding acquisition, N.M. All authors have read and agreed to the published version of the manuscript.}

\section*{Funding}{This research was funded by Office of Naval Research project Robot Aided Diver Navigation in Mapped Environments - ROADMAP under Grant Agreement No. N000142112274.}

\section*{Informed consent}{Informed consent was obtained from all subjects involved in the study.}

\section*{Data availability}{Datasets analyzed or generated during this study are publicly available at \url{https://labust.fer.hr/labust/research/datasets}.}

\section*{Acknowledgements}{Authors would like to thank all the diver participants of the experiment for their collaboration and patience. This work would not be possible without their enthusiasm and volunteering. They are, in alphabetical order: Allan Badian, Ante Nevistić, Christopher Walker, Derek Orbaugh, Helena Žigić, Igor Mašala, Kristijan Krčmar, Marta Plaza and Vedran Stipetić.}

\section*{Declaration of interests}{The authors declare that they have no known competing financial interests or personal relationships that could have appeared to influence the work reported in this paper.} 



  \bibliographystyle{elsarticle-num}


\begin{thebibliography}{10}
\expandafter\ifx\csname url\endcsname\relax
  \def\url#1{\texttt{#1}}\fi
\expandafter\ifx\csname urlprefix\endcsname\relax\def\urlprefix{URL }\fi
\expandafter\ifx\csname href\endcsname\relax
  \def\href#1#2{#2} \def\path#1{#1}\fi

\bibitem{uhri_overview}
A.~Birk, A survey of underwater human-robot interaction (u-hri), Current Robotics Reports 3 (2022) 1--13.
\newblock \href {https://doi.org/10.1007/s43154-022-00092-7} {\path{doi:10.1007/s43154-022-00092-7}}.

\bibitem{roadmap}
D.~Nađ, F.~Ferreira, I.~Kvasić, L.~Mandić, V.~Slošić, C.~Walker, D.~O. Antillon, I.~Anderson, Towards robot-aided diver navigation in mapped environments (roadmap), in: OCEANS 2022, Hampton Roads, 2022, pp. 1--5.
\newblock \href {https://doi.org/10.1109/OCEANS47191.2022.9977173} {\path{doi:10.1109/OCEANS47191.2022.9977173}}.

\bibitem{remote_experiments}
F.~Ferreira, I.~Kvasi\'{c}, D.~Na\dj, L.~Mandi\'{c}, N.~Mi\v{s}kovi\'{c}, C.~Walker, D.~O. Antillon, I.~Anderson, Diver‐robot communication using wearable sensing: Remote pool experiments, Marine technology society journal 56~(5) (2022) 26--35.
\newblock \href {https://doi.org/10.4031/mtsj.56.5.5} {\path{doi:10.4031/mtsj.56.5.5}}.

\bibitem{diver_assistance}
K.~J. DeMarco, M.~E. West, A.~M. Howard, Autonomous robot-diver assistance through joint intention theory, in: 2014 Oceans - St. John's, 2014, pp. 1--5.
\newblock \href {https://doi.org/10.1109/OCEANS.2014.7003003} {\path{doi:10.1109/OCEANS.2014.7003003}}.

\bibitem{safetyreq}
N.~Mišković, M.~Egi, D.~Nad, A.~Pascoal, L.~Sebastiao, M.~Bibuli, Human-robot interaction underwater: Communication and safety requirements, in: 2016 IEEE Third Underwater Communications and Networking Conference (UComms), 2016, pp. 1--5.
\newblock \href {https://doi.org/10.1109/UComms.2016.7583471} {\path{doi:10.1109/UComms.2016.7583471}}.

\bibitem{human_factor}
G.~Ho, N.~Pavlovic, R.~Arrabito, R.~Abdalla, Human factors issues when operating unmanned underwater vehicles, Proceedings of the Human Factors and Ergonomics Society Annual Meeting 55 (09 2011).
\newblock \href {https://doi.org/10.1177/1071181311551088} {\path{doi:10.1177/1071181311551088}}.

\bibitem{auvdivingrisks}
T.~Y. Loh, M.~P. Brito, N.~Bose, J.~Xu, K.~Tenekedjiev, A fuzzy-based risk assessment framework for autonomous underwater vehicle under-ice missions, Risk Analysis 39~(12) (2019) 2744--2765.
\newblock \href {https://doi.org/https://doi.org/10.1111/risa.13376} {\path{doi:https://doi.org/10.1111/risa.13376}}.

\bibitem{underwater_communication}
A.~N. Jaafar, H.~Ja'afar, I.~Pasya, R.~Abdullah, Y.~Yamada, Overview of underwater communication technology, in: K.~Isa, Z.~Md.~Zain, R.~Mohd-Mokhtar, M.~Mat~Noh, Z.~H. Ismail, A.~A. Yusof, A.~F. Mohamad~Ayob, S.~S. Azhar~Ali, H.~Abdul~Kadir (Eds.), Proceedings of the 12th National Technical Seminar on Unmanned System Technology 2020, Springer Singapore, Singapore, 2022, pp. 93--104.

\bibitem{acoustic_comms}
L.~Bjørnø, Chapter 14 - underwater acoustic measurements and their applications, in: T.~H. Neighbors, D.~Bradley (Eds.), Applied Underwater Acoustics, Elsevier, 2017, pp. 889--947.
\newblock \href {https://doi.org/https://doi.org/10.1016/B978-0-12-811240-3.00014-X} {\path{doi:https://doi.org/10.1016/B978-0-12-811240-3.00014-X}}.

\bibitem{WRIGHT1995}
J.~Wright, A.~Colling, Chapter 5 - light and sound in seawater, in: J.~Wright, A.~Colling (Eds.), Seawater: its Composition, Properties and Behaviour (Second Edition), second edition Edition, Pergamon, 1995, pp. 61--84.

\bibitem{gesture_communication}
H.~Buelow, A.~Birk, Gesture-recognition as basis for a human robot interface (hri) on a auv, in: OCEANS'11 MTS/IEEE KONA, 2011, pp. 1--9.
\newblock \href {https://doi.org/10.23919/OCEANS.2011.6107118} {\path{doi:10.23919/OCEANS.2011.6107118}}.

\bibitem{caddy-general}
N.~Mišković, A.~Pascoal, M.~Bibuli, M.~Caccia, J.~A. Neasham, A.~Birk, M.~Egi, K.~Grammer, A.~Marroni, A.~Vasilijević, Z.~Vukić, Caddy project, year 1: Overview of technological developments and cooperative behaviours, IFAC-PapersOnLine 48~(2) (2015) 125--130, 4th IFAC Workshop onNavigation, Guidance and Controlof Underwater VehiclesNGCUV 2015.
\newblock \href {https://doi.org/https://doi.org/10.1016/j.ifacol.2015.06.020} {\path{doi:https://doi.org/10.1016/j.ifacol.2015.06.020}}.

\bibitem{caddian}
D.~Chiarella, M.~Bibuli, G.~Bruzzone, M.~Caccia, A.~Ranieri, E.~Zereik, L.~Marconi, P.~Cutugno, Gesture-based language for diver-robot underwater interaction, in: OCEANS 2015 - Genova, 2015, pp. 1--9.
\newblock \href {https://doi.org/10.1109/OCEANS-Genova.2015.7271710} {\path{doi:10.1109/OCEANS-Genova.2015.7271710}}.

\bibitem{padi}
{Denny, M, PADI Blog - Scuba Diving and Freediving Tips, Dive Travel Insights}, Scuba diving hand signals, \url{https://blog.padi.com/scuba-diving-hand-signals}, accessed: 2023-9-06 (Jul 2022).

\bibitem{CMAS}
{Recreational Scuba Training Council, NEADC online PDF}, Common hand signals for scuba diving, \url{http://www.neadc.org/CommonHandSignalsforScubaDiving.pdf}, accessed: 2023-10-11 (2005).

\bibitem{caddian2}
D.~Chiarella, M.~Bibuli, G.~Bruzzone, M.~Caccia, A.~Ranieri, E.~Zereik, L.~Marconi, P.~Cutugno, \href{https://www.mdpi.com/2077-1312/6/3/91}{A novel gesture-based language for underwater human–robot interaction}, Journal of Marine Science and Engineering 6~(3) (2018).
\newblock \href {https://doi.org/10.3390/jmse6030091} {\path{doi:10.3390/jmse6030091}}.
\newline\urlprefix\url{https://www.mdpi.com/2077-1312/6/3/91}

\bibitem{CADDY-dataset}
A.~Gomez~Chavez, A.~Ranieri, D.~Chiarella, E.~Zereik, A.~Babić, A.~Birk, \href{https://www.mdpi.com/2077-1312/7/1/16}{Caddy underwater stereo-vision dataset for human–robot interaction (hri) in the context of diver activities}, Journal of Marine Science and Engineering 7~(1) (2019).
\newblock \href {https://doi.org/10.3390/jmse7010016} {\path{doi:10.3390/jmse7010016}}.
\newline\urlprefix\url{https://www.mdpi.com/2077-1312/7/1/16}

\bibitem{underwater_image_processing}
H.~Lu, Y.~Li, Y.~Zhang, M.~Chen, S.~Serikawa, H.~Kim, \href{http://arxiv.org/abs/1702.03600}{Underwater optical image processing: {A} comprehensive review}, CoRR abs/1702.03600 (2017).
\newblock \href {http://arxiv.org/abs/1702.03600} {\path{arXiv:1702.03600}}.
\newline\urlprefix\url{http://arxiv.org/abs/1702.03600}

\bibitem{glove}
D.~W.~O. Antillon, C.~R. Walker, S.~Rosset, I.~A. Anderson, Glove-based hand gesture recognition for diver communication, IEEE Transactions on Neural Networks and Learning Systems (2022) 1--13\href {https://doi.org/10.1109/TNNLS.2022.3161682} {\path{doi:10.1109/TNNLS.2022.3161682}}.

\bibitem{elastomers}
D.~W.~O. Antillon, C.~Walker, S.~Rosset, I.~A. Anderson, \href{https://doi.org/10.1117/12.2558388}{{The challenges of hand gesture recognition using dielectric elastomer sensors}}, in: Y.~Bar-Cohen (Ed.), Electroactive Polymer Actuators and Devices (EAPAD) XXII, Vol. 11375, International Society for Optics and Photonics, SPIE, 2020, p. 1137524.
\newblock \href {https://doi.org/10.1117/12.2558388} {\path{doi:10.1117/12.2558388}}.
\newline\urlprefix\url{https://doi.org/10.1117/12.2558388}

\bibitem{swimpro}
Swimpro, Underwater cameras for swimmers, \url{https://swimpro.com.au/}, accessed: 2023-10-09 (2023).

\bibitem{unity}
A.~Juliani, V.-P. Berges, E.~Teng, A.~Cohen, J.~Harper, C.~Elion, C.~Goy, Y.~Gao, H.~Henry, M.~Mattar, D.~Lange, Unity: A general platform for intelligent agents (2020).
\newblock \href {http://arxiv.org/abs/1809.02627} {\path{arXiv:1809.02627}}.

\bibitem{gestures}
M.~J. Islam, \href{http://arxiv.org/abs/1804.02479}{Understanding human motion and gestures for underwater human-robot collaboration}, CoRR abs/1804.02479 (2018).
\newblock \href {http://arxiv.org/abs/1804.02479} {\path{arXiv:1804.02479}}.
\newline\urlprefix\url{http://arxiv.org/abs/1804.02479}

\bibitem{nanomodem}
B.~Sherlock, N.~Morozs, J.~Neasham, P.~Mitchell, \href{https://www.mdpi.com/2079-9292/11/9/1446}{Ultra-low-cost and ultra-low-power, miniature acoustic modems using multipath tolerant spread-spectrum techniques}, Electronics 11~(9) (2022).
\newblock \href {https://doi.org/10.3390/electronics11091446} {\path{doi:10.3390/electronics11091446}}.
\newline\urlprefix\url{https://www.mdpi.com/2079-9292/11/9/1446}

\bibitem{multipath}
J.~M. Hovem, H.~Dong, \href{https://www.mdpi.com/2077-1312/7/4/118}{Understanding ocean acoustics by eigenray analysis}, Journal of Marine Science and Engineering 7~(4) (2019).
\newblock \href {https://doi.org/10.3390/jmse7040118} {\path{doi:10.3390/jmse7040118}}.
\newline\urlprefix\url{https://www.mdpi.com/2077-1312/7/4/118}

\end{thebibliography}






\end{document}